\documentclass[%
 preprint,
 superscriptaddress,
 amsmath,amssymb,
 aps,
 floatfix,
]{revtex4-2}

\usepackage{graphicx}
\usepackage{dcolumn}
\usepackage{bm}
\usepackage{hyperref}
\usepackage{palatino,euler}

\begin{document}
\title{Transforming Design Spaces Using Pareto-Laplace Filters}

\author{Hazhir Aliahmadi}
\affiliation{Department of Physics, Engineering Physics,
and Astronomy, Queen's University, Kingston ON, K7L 3N6, Canada}
\author{Ruben Perez}
\affiliation{Department of Mechanical and Aerospace Engineering, Royal Military
  College of Canada, Kingston ON, K7K 7B4}
\author{Greg van Anders}
\affiliation{Department of Physics, Engineering Physics,
and Astronomy, Queen's University, Kingston ON, K7L 3N6, Canada}
\email{gva@queensu.ca}

\date{\today}

\begin{abstract}
  Optimization is a critical tool for addressing a broad range of human and technical problems. However, the paradox of advanced optimization techniques is that they have maximum utility for problems in which the relationship between the structure of the problem and the ultimate solution is the most obscure. The existence of solution with limited insight contrasts with techniques that have been developed for a broad range of engineering problems where integral transform techniques yield solutions and insight in tandem. Here, we present a ``Pareto-Laplace'' integral transform framework that can be applied to problems typically studied via optimization. We show that the framework admits related geometric, statistical, and physical representations that provide new forms of insight into relationships between objectives and outcomes. We argue that some known approaches are special cases of this framework, and point to a broad range of problems for further application.
\end{abstract}

\maketitle

\section{Introduction}
Integral transforms serve an indispensable function in a broad range of
engineering problems\cite{AdvEngMath}. For example in signal processing
\cite{oppenheimSignalsSystems1983} and control theory,\cite{ogataModernControlEngineering2010} although the real world phenomena of
interest play out in time, the structure of engineering systems and the design
of systems is done in the frequency domain.\cite{AdvEngMath}  Similarly, in
various engineering domains where behaviours are governed by waves, integral
transforms can render more interpretable illustrations of the phenomena that
facilitate engineering design\cite{debnathIntegralTransformsTheir2007}. The plurality of domains in which
integral transforms are critical for engineering design raises the question of
whether analogous approaches could provide similar levels of utility in other
areas of engineering and design where they are not currently part of the standard
practice.

The need for more sophisticated forms of analysis and understanding
is particularly pressing in engineering domains that make
extensive use of optimization. Optimization problems are notoriously difficult
to solve \cite{natureofcomputation} and solution algorithms frequently rely on
techniques that obscure the underlying structure of the problem
\cite{martinsMultidisciplinaryDesignOptimization2013}. Although the situation is
different in some case, for example gradient-based techniques (see, e.g., Ref.\
\cite{PolakOptimization}) yield information about
the local structure of the solution space, comprehensive, global pictures are
more difficult to construct. The lack of
clear relationships between the problem structure and the solution characteristics
for many optimization problems raises a number of issues.
Those issues include questions about sensitivity and future adaptability, among
others. 

Here, we describe an integral transform framework that can be applied to
problems that are conventionally studied via optimization. The framework employs
one or more objective functions to foliate the solution space to a design
problem in terms of Pareto surfaces, and then applies a Laplace transform to the
generalized volume of the surface. This ``Pareto-Laplace'' framework effectively
filters the solution space by exponentially suppressing regions with large
objective function values. As we show below, key features of the solution space
yield identifiable effects in the Pareto-Laplace filtered form of the problem.

A key feature of the Pareto-Laplace framework is that it can be cast in
equivalent geometric, statistical, and physical representations, which open
avenues for powerful insight into the structure of underlying design problems.
Geometrically, the Pareto-Laplace framework represents a filter on the solution
space, where the Laplace variable scales the contributions from solutions based
on their objective function values, analogous to a hyperbolic projection.
Statistically, the Laplace transform serves as a moment-generating function 
(see, e.g., Ref.\ \cite{shannon}),
allowing for a probabilistic interpretation of the solution space volume, which
can be computed without explicit knowledge of key quantities such as the minimum
objective value or the volume function. Physically, it takes the form of a
partition function in statistical mechanics (see, e.g., Ref.\
\cite{Sethna2021}), relating to thermodynamic concepts such as temperature and
energy.

A second key feature of the framework is that its tripartite
geometric/statistical/physical representation yields powerful computational
approaches. The existence of means to explore solution spaces, e.g., via
techniques including Monte Carlo and molecular dynamics simulation, means that
it is possible to implement the filter on complex problems where the quantities
that enter the formal definition of the filter are unknown.

A third key feature of the Pareto-Laplace filtration framework is that it is
that related approaches to design and optimization problems emerge as special
cases in certain limits. Special cases include simulated annealing
\cite{kirkpatrick1983}, as well as statistical mechanics based materials
design techniques such as digital alchemy \cite{digitalalchemy} and
gradient-based approaches.\cite{miskinpnas}

The remainder of this paper describes the structure of the
Pareto-Laplace filter. We motivate the filter based on general arguments, and
then interpret the resulting mathematical quantities in geometric, statistical,
and physical terms. We derive the effect of problem reparameterization on the
structure of the filter. To build intuition about the structure of the framework
we give explicit results for an example in linear programming where it is
possible to compute the framework in closed form. Our primary interest is in
applications of the filter in nonlinear and non-convex optimization problems,
and example applications of the framework in problems ranging from naval
architecture \cite{nfl}, land-use planning,\cite{Flashpoints,slo-go}, as
well prior applications in materials design,\cite{digitalalchemy,engent} can be
found in other works.

\section{Pareto-Laplace Filter}
In this section, we motivate and derive the Pareto-Laplace filter from general
arguments. We then give a trio of representations of the resulting mathematical
expressions.

For illustrative purposes and to keep the development self-contained, we work
first with a single objective function without constraints, before generalizing
to an arbitrary number of objectives and/or constraints. Moreover, for the
purposes of maintaining readability we avoid weighing down the description with
rigorous proofs.

\subsection{Motivation and Derivation} 
Consider an optimization problem
\begin{equation}
  \min_{x\in\mathcal{S}} \mathcal{O}(x) \;
  \label{eq:OptProb}
\end{equation}
i.e., the minimization of some objective function $\mathcal{O}$ over some set of
possible design solutions $\{x\}$ in a solution space $\mathcal{S}$. A typical
optimization algorithm aims to identify the solution of the problem $x_*$ and
determine $\mathcal{O}_\text{min}\equiv\mathcal{O}(x_*)$, the objective at the
minimum. However, this solution may not provide an answer to all design
questions related to $\mathcal{O}$ and $\mathcal{S}$.

Many issues, e.g., sensitivity analysis or algorithmic implementation, depend
not only on the solution, but on the structure of the solution space. In
general, for a given optimization problem, neither the optimal solution nor the
structure of the solution space is known.  To determine the structure of the
space, since we are interested in the space in the context of a specific
optimization objective it strongly suggests we foliate the solution space by
slices through the space for which $\mathcal{O}$ is constant. We will refer to
this as a Pareto slicing because surfaces or level sets in $\mathcal{O}$ play a
central role in Pareto-style approaches to optimization. Since the solution
space must be geometrical in some sense, the basic notions of the structure of
the space should be encoded by some geometric measure on the space. A very basic
geometric measure is given by its volume. We are particularly interested in the
volume of the Pareto slices that lie in a direction that is transverse to
$\mathcal{O}$, so we will represent the volume of the solution space $\Omega$ as
a function of the objective function
\begin{equation}
  \Omega = \int d\mathcal{O} \Omega_\perp(\mathcal{O}) \; .
  \label{eq:OmegaDef}
\end{equation}

Armed with a primitive geometric measure of the solution space, it raises the
problem of how to formulate a meaningful integral transform. In optimization,
near-optimal solutions are likely to yield more useful information than highly
sub-optimal ones. I.e., for many questions ``good'' solutions could be more
instructive than ``bad'' ones, though the degree to which this is the case may
vary (e.g., what counts as good or bad could be context dependent). Indeed, a
systematic way of varying the tolerance for non-optimal solutions would provide
an entry point for sensitivity analysis. Moreover, we should expect that in
generic optimization problems, in the absence of constraints to the contrary,
there will likely be far more ``bad'' potential solutions than ``good'' ones.
Therefore one should expect $\Omega_\perp$ will grow with $\mathcal{O}$ in
generic cases. Anticipating $\Omega_\perp$ grows with $\mathcal{O}$, a functional
filter should strongly suppress large-$\mathcal{O}$ regions of $\mathcal{S}$. 

Considering expectations of the relative value of ``good'' and ``bad''
solutions and together with the growth of $\Omega_\perp$ with $\mathcal{O}$
suggests filtering the solution space via a Laplace transform of the form
\begin{equation}
  Z(\beta) =
  \int_{\mathcal{O}_\text{min}(\mathcal{C})}^{\infty}d\mathcal{O} e^{-\beta
  \mathcal{O}} \Omega_\perp(\mathcal{O}) \; , \label{eq:LTfamiliar}
\end{equation}
where $\beta$ plays the role of the Laplace variable, and
$\mathcal{O}_\text{min}$ is the minimum possible value for the objective. Note
Eq.\ \eqref{eq:LTfamiliar} satisfies the criteria we outlined: the parameter
$\beta>0$ controls the relative contribution to $Z$ from regions of
$\mathcal{S}$ with low- or high values of $\mathcal{O}$, and it exponentially
suppresses solutions with large $\mathcal{O}$.

\subsection{Geometric, Statistical, and Physical Representations}
\label{sec:3reps}
Eq.\ \eqref{eq:LTfamiliar} gives a formal definition of the Pareto-Laplace
filter. However, the key to the utility of the filter lies in the interpretation
of $Z(\beta)$. In this subsection, we first consider the geometric aspects of the
filter in more detail. We then find that this geometric representation suggests
alternate statistical and physical representations.
This trio of representations is a key to the Pareto-Laplace filter's versatility
and its use in illuminating how the solution space of a design problem
is influenced by the design objectives (see Sec.\ \ref{sec:Analysis}).

\subsubsection{Geometric Representation} \label{sec:GeoRep}
\begin{figure}
  \begin{center}
  \includegraphics{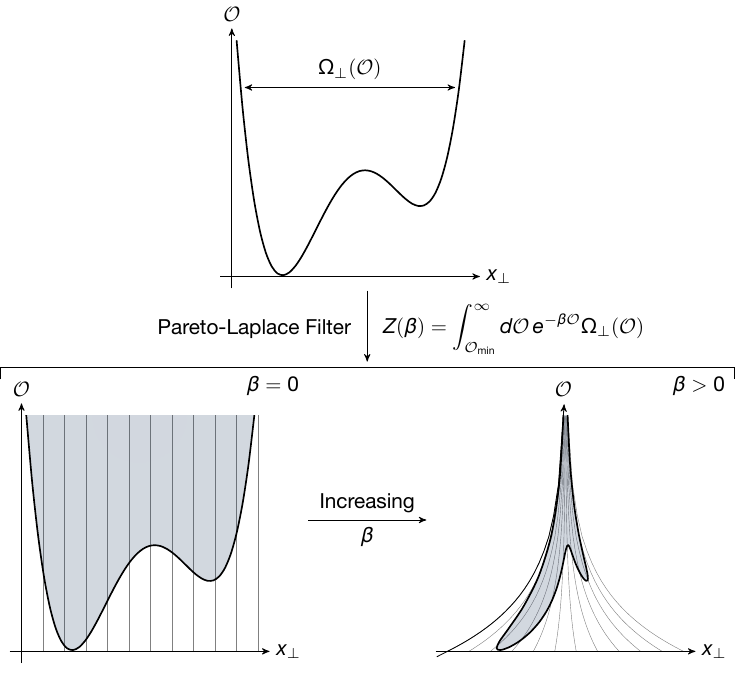}
  \end{center}
  \caption{Schematic representation of design landscape integral transform for a
    single design objective. (Top middle) A design problem exhibits a landscape of
    potential solutions distributed over a solution space $\mathcal{S}$ in coordinates
    $(\mathcal{O},x_\perp)$ according to the level at which they satisfy the
    design objective $\mathcal{O}$ and a set of other design variables $x_\perp$
    (represented here schematically as a single variable).  There may be
    $\Omega_\perp$ solutions that satisfy the design objective at some fixed
    level $\mathcal{O}$. We propose to filter the solution space via
    the application of a integral transform, where $Z$ computes the volume of
    the solution space depending a parameter $\beta$ that controls the degree of
    filtering of poor design solutions (which we take as large $\mathcal{O}$).
    For $\beta=0$ (lower left) there is no filtering, however increasing $\beta$
    (lower right) effectively ``pinches'' the landscape for
    large $\mathcal{O}$. Increasingly large values of $\beta$ leave effectively
    larger relative contributions from near optimal solutions, with only the
    optimal solution remaining in the limit $\beta\to\infty$.
  }
  \label{fig:hyperbolic}
\end{figure}

Since $\Omega(\mathcal{O})$ encodes the ``volume'' of potential
design solutions that realize the design objective $\mathcal{O}$ at some fixed
level, the quantity $Z(\beta)$ aggregates those volumes to give a total volume
of the solution space for all possible $\mathcal{O}$, however the contributions
from solutions are scaled by a factor $\exp(-\beta\mathcal{O})$. In particular,
a single solution at $\mathcal{O}$ is suppressed relative to a solution at
$\mathcal{O}_\text{min}$ by $w(\mathcal{O}) = e^{-\beta(\mathcal{O}-\mathcal{O}_\text{min})}$.
We sketch this schematically in Fig.\ \ref{fig:hyperbolic} (see also Supplementary
Movies S1 and S2).

We arrived at Eq.\ \eqref{eq:LTfamiliar} by starting with an optimization
defined over a solution space. Let us call the solution space $\mathcal{S}$. If
we had $N$ real, continuous design variables then we would have that
$\mathcal{S}\subset\mathbb{R}^N$. If we take the objective function $\mathcal{O}$ as a
mapping $\mathcal{O}:\mathcal{S}\to \mathbb{R}$, then we define the solution landscape as
a foliation of $(\mathcal{O},x_\perp)\in \mathcal{S}$ where $x_\perp$ lives on the
$(N-1)$-dimensional space of foliations that are perpendicular to $\mathcal{O}$
in $\mathcal{S}$.

Momentarily considering the specific case of linear programming provides valuable
intuition. Considering a case where $\mathcal{O}=-\vec{C}\cdot\vec{x}$, in our
geometric picture $\mathcal{O}$ corresponds to the $\vec{C}$ direction in
$\mathcal{S}$, $\vec{C}\cdot\vec{x}_\perp=0$, and $\Omega_\perp(\mathcal{O})$
corresponds to the volume of the feasible region along the Pareto fronts.

Returning to the general problem, as shown in Fig.\ \ref{fig:hyperbolic}
geometrically, Eq.\ \eqref{eq:LTfamiliar} implements a filter on the solution
space. It is possible to write a metric in the filtered geometry, which we will
refer to as $\mathcal{S}_\beta$, as
\begin{equation}
  ds^2 = d\mathcal{O}^2 + e^{-2\beta\mathcal{O}/(N-1)} ds_{\Omega_\perp}^2
  \label{eq:Metric}
\end{equation}
where $ds$ is the line element in the filtered space, $N$ is the number of
dimensions in the solution space, and $ds_{\Omega_\perp}$ is the induced line
element in the solution space that is orthogonal to $\mathcal{O}$. The geometric
nature of the filter is clarified by considering the coordinate transformation
$a = \tfrac{N-1}{\beta}\exp(\beta\mathcal{O}/(N-1))$, which gives the metric
as
\begin{equation}
  ds^2 =\frac{da^2+ds_{\Omega_\perp}^2}{a^2} \; ,
  \label{eq:Hyp2Metric}
\end{equation}
which is a standard form for the metric of an $N$-dimensional hyperbolic space if
the metric on the transverse space is flat, i.e.\
$ds_{\Omega_\perp}^2=d\vec{x}_\perp^2$.

In light of Eq.\ \eqref{eq:Metric}, one can formally recover Eq.\
\eqref{eq:LTfamiliar} by computing the volume measure on the filtered space,
\begin{equation}
  d\omega = d\mathcal{O} d^{N-1}x_\perp e^{-\beta \mathcal{O}} \; ,
  \label{eq:domega}
\end{equation}
and computing the volume $Z(\beta)=\int d\omega$ by integrating the solution
space volume via foliation at a series of fixed $\mathcal{O}$.

Further technical geometric details relating the behavior of the objective
function to the geometry of the solution space can be found in Appendices
\ref{ap:multi} and \ref{ap:decrease}, and on geodesics in $\mathcal{S}_\beta$
can be found in Appendix \ref{ap:Geodesic}. 

\subsubsection{Statistical Representation} \label{sec:StatRep}
As we will show below, it is possible to explicitly compute $Z(\beta)$ for some
problems, such as those in linear programming, explicit, closed-form evaluation
is not feasible for many nonlinear and non-convex problems. In most optimization
problems, we do not expect to have direct, a priori knowledge of
$\mathcal{O}_\text{min}$. Moreover, we do not expect to know
$\Omega_\perp(\mathcal{O})$ explicitly. Given the lack of explicit knowledge of
any of the key quantities that appear in Eq.\ \eqref{eq:LTfamiliar} one might
expect that, whatever its potential use, computing $Z(\beta)$ is inherently
fraught.

However, we will note that the essence of the problem of computing Eq.\
\eqref{eq:LTfamiliar} is to compute a volume. Computing volumes where direct
integration is impractical is a standard problem in computational science, and
there are many approaches to solving integration problems by recasting them in
statistical language. E.g., it is a standard textbook exercise to use what is
sometimes referred to as the ``hit or miss'' method (see, e.g.\ Ref.\
\cite{GouldTobochnik3e})
to to estimate $\pi$ by Monte Carlo integration.

Although basic algorithmic considerations could lead to a statistical
representation of Eq.\ \eqref{eq:LTfamiliar} it is more instructive to arrive at
that perspective starting from an information theoretic point of view.
Information theory gives a recipe for constructing probability distributions
with a given set of moments that make no other assumptions about the form of the
distribution. That distribution is found by maximizing the entropy $S$ with
respect a probability distribution $p(x)$,
\begin{equation}
  S = \int_{\mathcal{S}} d^Nx p(x) \ln(p(x))
  +\beta\left(\int_{\mathcal{S}} d^Nx p(x) \mathcal{O}(x) - \left<\mathcal{O}\right>\right)
  -\lambda\left(\int_{\mathcal{S}} d^Nx p(x) -1\right) \; ,
  \label{eq:entropy}
\end{equation}
where $\beta$ is a Lagrange multiplier that enforces the moment constraint and
$\lambda$ is a Lagrange multiplier that enforces normalization. The maximization
of Eq.\ \eqref{eq:entropy} gives
\begin{equation}
  p(x) = \frac{e^{-\beta\mathcal{O}}}{Z(\beta)} \; ,
\end{equation}
where
\begin{equation}
  Z(\beta) = \int_{\mathcal{S}} d^Nx e^{-\beta\mathcal{O}} \; .
  \label{eq:ZintIT}
\end{equation}
Now, if we compute the integral by decomposing it as
$d^Nx=d\mathcal{O}d^{N-1}x_\perp$ and computing the integral over $x_\perp$, we
obtain
\begin{equation}
  Z(\beta) = \int_{\mathcal{O}_\text{min}}^{\infty} d\mathcal{O} \,
  e^{-\beta\mathcal{O}} \Omega_\perp(\mathcal{O}) \; .
  \label{eq:LTfamiliarRedux}
\end{equation}
Note that this is precisely the same form as Eq.\ \eqref{eq:LTfamiliar} if we
identify $\beta$, the Lagrange multiplier in Eq.\ \eqref{eq:LTfamiliarRedux},
with $\beta$, the Laplace variable Eq.\ \eqref{eq:LTfamiliar}.

The connection between the geometric integral transform of optimization in Eq.\
\eqref{eq:LTfamiliar} and the statistical interpretation of that quantity is
not coincidental. In fact, for design problems in other areas of science and
engineering, e.g., the design of self-assembled
materials,\cite{digitalalchemy,engent,packingassembly,Alch-MD,bccfccdesign,identitycrisis}
naval architecture,\cite{systemphys,robustdesign,aaa} and land-use planning,
\cite{Flashpoints,slo-go} partition functions of the form Eq.\ \eqref{eq:Zdef}
have been derived starting not from a geometric point of view, but from an
information theoretic one.\cite{shannon,jaynes1}

\subsubsection{Physical Representation}
The deep connection between information theory and statistical mechanics
\cite{jaynes1} suggests the existence of a third, physical representation of the
Pareto-Laplace filter in Eq.\ \eqref{eq:LTfamiliar}. $Z(\beta)$ can be
identified as a partition function in statistical physics where $\beta=1/T$,
where $T$ is thermodynamic temperature, $\mathcal{O}$ plays the role of energy,
$\mathcal{O}_\text{min}$ plays the role of the ground state energy, and
$\Omega(\mathcal{O})$ plays the role of the density of states, e.g.\ refer to
Ref.\ \cite{Sethna2021}.

The existence of a physical representation of the Pareto-Laplace filter has two
important implications: one practical and one conceptual.

From the practical point of view, physics has developed a broad array of
well-defined computational techniques, see, e.g.\ Ref.\ \cite{frenkel}, that
provide statistical sampling that can generate the distributions of $\{x\}$ that
contribute to $Z(\beta)$ up to arbitrary accuracy even without a priori
knowledge of the ground state energy and the density of states.
The key to implementing these approaches is to recast Eq.\ \eqref{eq:LTfamiliar}
as
\begin{equation}
  Z(\beta) = \int_{\mathcal{S}} d^{N}x e^{-\beta\mathcal{O}(x)}
  \label{eq:Zdef}
\end{equation}
where one does not assume that one has a priori knowledge of how to foliate
$\mathcal{S}$ in terms of $\mathcal{O}$. Given this framing, a very wide range
of techniques can be used to sample the distributions that generate $Z(\beta)$
via techniques such as Monte Carlo (MC) and molecular dynamics (MD) simulation
depending on the form of $\mathcal{S}$ and $\mathcal{O}(x)$.

From the conceptual point of view, the fact that it is possible to impart Eq.\
\eqref{eq:LTfamiliar} with a physical representation in terms of thermodynamics
points to one of the advantages of thermodynamics as a description of systems of
many degrees of freedom: it abstracts the complex interplay of the many degrees
of freedom in terms of concrete, tangible mechanical effects such as pressure,
stress, strain, etc. Aspects of this perspective have been developed starting
from the statistical representation of this framework in materials design under
what the authors of Ref.\ \cite{digitalalchemy} termed ``digital alchemy''
and in the context of networks under what the authors of \cite{systemphys}
termed ``systems physics''.

\subsection{Symmetry and Problem Reformulation}
\label{sec:symmetry}
The solution space filter Eq.\ \eqref{eq:LTfamiliar} provides considerable
information about the structure of the underlying design problem. However, for
many problems, it is possible to formulate them in multiple ways (e.g., via
different choices of parameters or units), and it is therefore crucial to
understand how such problem reformulations affect the information that the
transform yields.

\subsubsection{Translation}
One way of reformulating the design problem would be to shift coordinates in
$\mathcal{S}$ via a constant. Two cases are instructive.

If the coordinate transformation has the form
\begin{equation}
  x_\perp \to x_\perp + \Delta x_\perp \qquad
  \mathcal{O}\to\mathcal{O} \; ,
  \label{eq:xshift}
\end{equation}
i.e., there is a shift in the transverse directions that leaves the objective
unchanged, then in Eq.\ \eqref{eq:LTfamiliar} the measure and the scaling factor
are invariant, and $\Omega_\perp \to \Omega_\perp$ so it is also invariant. As a
result, $Z(\beta)\to Z(\beta)$, which means that Eq.\ \eqref{eq:LTfamiliar}
preserves all information under transverse translation.

If the coordinate transformation has the form
\begin{equation}
  x_\perp \to x_\perp \qquad
  \mathcal{O}\to\mathcal{O} + \Delta\mathcal{O} \; ,
  \label{eq:Oshift}
\end{equation}
i.e., there is a shift in the objective that leaves the transverse directions
unchanged, then in Eq.\ \eqref{eq:LTfamiliar} the measure
and $\Omega_\perp \to \Omega_\perp$ are invariant, however with the scaling factor
$e^{-\beta\Delta\mathcal{O}}$. As a
result, $Z(\beta)\to e^{-\beta\Delta\mathcal{O}} Z(\beta)$. The effect of this
scaling factor is to shift appropriate moments of $\mathcal{O}$ by
$\Delta\mathcal{O}$, however, it preserves the rest of the structure of
$Z(\beta)$.

\subsubsection{Rotation}
In addition to translation, consider rotation in $\mathcal{S}$. Rotations that
correspond to coordinate transformations on $\mathcal{S}$ that do not alter the
objective function preserve $\Omega_\perp$, the measure, and scaling factor in
the transform. Hence $Z(\beta)\to Z(\beta)$, so the transform preserves
information under rotations that do not alter the objective.

\subsubsection{Rescaling}
Finally, it is instructive to consider rescaling the objective. Objective
rescaling could arise if, say, an objective was reformulated in different units. 

Consider $\mathcal{O}\to\alpha\mathcal{O}$. This transformation preserves
$\Omega_\perp$, however, the measure and scaling factor are not invariant.
However, we not that applying this rescaling in Eq.\ \eqref{eq:LTfamiliar} gives
$Z(\beta) \to \frac{1}{\alpha}Z(\beta/\alpha)$. That means that up to a constant
overall rescaling of $Z(\beta)$ and a rescaling $\beta\to\beta/\alpha$
the transform
preserves all information about the design space.

\subsection{Generalizations}
For simplicity we developed the theory above for problems with a single
objective and no constraints and a continuous $\mathcal{S}$. In this section we
relax these assumptions.
\subsubsection{Multi-Objective Problems}
We are interested in problems where we can have multiple objectives. Given the
multiple representations of the design filter that we identified in the case of
a single objective function, there are several equivalent routes one could take
that to develop an analogous theory for multiple design objectives. We will give
one line of argument that requires a minimal amount of mathematical formalism
and leave the description of derivations via other approaches to other work.

For those multi-objective problems it is useful to foliate the solution space
$\mathcal{S}$ according to set of $M$ design objectives $\mathcal{O}_i$, where
we are interested in the volume of the solution space where each of the
$\mathcal{O}_i$ is fixed $\Omega_\perp(\{\mathcal{O}_i\})$. In principle it
would be useful to implement a filter of $\Omega$ according to each of the
design objectives. We therefore write
\begin{equation}
  Z(\{\beta_i\}) = \left[\prod_{i=1}^M \int_{\mathcal{O}^{(i)}_\text{min}}^\infty
  d\mathcal{O}_i e^{-\beta_i \mathcal{O}_i}\right]
  \Omega_\perp(\{\mathcal{O}_i\}) \; .
  \label{eq:LTmulti}
\end{equation}

We can interpret $Z(\{\beta_i\})$ as a volume on the space with the line element
\begin{equation}
  ds^2 = \sum_{i=1}^M d\mathcal{O}_i^2
  e^{-\frac{2}{N-1}\sum_{j\neq i} \beta_j\mathcal{O}_j}
  +e^{-\frac{2M}{N-1}\sum_{i=1}^M \beta_i\mathcal{O}_i}dx_\perp^2 \; ,
  \label{eq:ds2multi}
\end{equation}
which gives the volume element
\begin{equation}
  d\omega =\left[\prod_{i=1}^M d\mathcal{O}_i e^{-\beta_i \mathcal{O}_i}\right]
  d^{N-M}x_\perp \; .
  \label{eq:omegaMulti}
\end{equation}
One can recover Eq.\ \eqref{eq:LTmulti} by integrating $d\omega$ over
$\mathcal{S}$ by integrating each slice of $x_\perp$.

Note that, like the single objective case, it is a straightforward exercise in
information theory to derive Eq.\ \eqref{eq:LTmulti} via entropy maximization.

It is useful to remark briefly on the interpretation of various quantities that
are computable from Eq.\ \eqref{eq:LTmulti}. It will not be surprising since the
single objective Eq.\ \eqref{eq:LTfamiliar} admitted geometrical, statistical,
and physical interpretations, the multi-objective case yields a similar
range of interpretations. Some multi-objective applications of this framework
have been described in other work, e.g.\ Refs.\
\cite{bccfccdesign,systemphys,Flashpoints,slo-go}.

It is important to note that although we are considering examples in which there
are multiple objectives at play, the focus of our analysis was on the form of
the solution space as ``scored'' by the various objectives of interest. However,
it is important to note than in the treatment of the multi-objective case, at no
point in the analysis were we forced to write a single overall objective
function. In this sense, our analysis is more general than optimization and is
not predicated on the choice of any one particular form for the optimization
problem. It is therefore possible to situate many different formulations
optimization problems involving the same set of objectives in the context of
this framework. Indeed, that could be regarded as one of the strengths of the
present approach.
\subsubsection{Constraints}
Until this point, we've avoided a detailed discussion of constraints, however
they are straightforward to incorporate into this framework.  For inequality
constraints, one must simply cast constraints in the form
\begin{equation}
  g_\gamma(x) > 0
  \label{eq:ineqc}
\end{equation}
and multiply the integrand in Eq.\ \eqref{eq:LTfamiliar} or Eq.\
\eqref{eq:LTmulti}
\begin{equation}
  G(x) = \prod_{\gamma} \theta(g_\gamma(x)) \; ,
  \label{eq:Gx}
\end{equation}
where $\theta(\cdot)$ is the unit step function. The case of equality
constraints can be handled by formulating them so that
\begin{equation}
  h_\delta(x) = 0
  \label{eq:equality}
\end{equation}
and multiplying the transform integrand by
\begin{equation}
  D(x) = \prod_{\delta} \delta(h_\delta(x))
  \label{eq:Dx}
\end{equation}
where $\delta(\cdot)$ is the Dirac delta function.

Readers may recognize that constraints could simply be absorbed into the
definition of $\mathcal{S}$, however in cases where solving the
constraints is difficult, it could be more profitable to define $\mathcal{S}$ as
unconstrained and write an explicit set of constraints. In this case,
notationally it would make sense to write the volume of the solution space
foliation as
\begin{equation}
  \Omega_\perp(\{\mathcal{O}_i\}) \to 
  \Omega_\perp(\{\mathcal{O}_i\},\{\kappa_\alpha\}) \; ,
  \label{eq:OmegaConstraint}
\end{equation}
where $\{\kappa_\alpha\}$ are some parameters describing the constraints.

Readers may also recognize there is nothing to stop one from using the transform
framework to study the structure of the solution space as a function of the
constraints alongside or instead of as a function of objectives.
Constraint-based transformations also yield relationships among moments that
provide leverage. Indeed, there are cases in which performing a transformation
with respect to a constraint provides very useful information about the
structure of the solution space; see, e.g., Refs.\
\cite{packingassembly,HyperoptMorphogenesis}.

Note, viewing the Pareto-Laplace framework from the physics perspective, it is also
straightforward to incorporate constraints using Lagrange multiplier methods. Using
Lagrange multipliers to invoke constraints can be implemented simply by performing
a Legendre transform (see, e.g., Ref.\ \cite{LLv5} for a detailed description of
Legendre transforms in conventional, statistical physics settings). This approach has been applied in
prior works that use the Pareto-Laplace framework
in specific cases, e.g.\ Refs.\ \cite{digitalalchemy,packingassembly,HyperoptMorphogenesis}.

\subsubsection{Discrete Cases}
Many problems of interest involve discrete spaces; e.g., instead of
$\mathcal{S}\in\mathbb{R}^N$ those problems have $\mathcal{S}\in\mathbb{Z}^N$.
The theory we described above carries over straightforwardly to that case. In
particular, Eq.\ \eqref{eq:LTfamiliar} retains its form, the only difference is
that $\Omega_\perp(\mathcal{O})$ rather than being a continuous volume is a sum of
delta-functions at each allowed discrete value of $\mathcal{O}$ where the
coefficient of the delta function is the discrete number of solutions in
$\mathcal{S}$ with that value of $\mathcal{O}$.

Notions we developed in the continuum about, e.g., moments, including their
geometric, statistical, and physical interpretation all carry over to the
discrete case. The only difference is that in the discrete case, these notions
refer to sets of points in space rather than a continuum volume.

Continuum intuition about modes in $Z(\beta)$ that arise collectively from dense
regions in $\mathcal{S}_\beta$ also have meaning in discrete cases if one interprets
density in $\mathcal{S}_\beta$ as a smeared-out version of density over a region of
$\mathcal{S}_\beta$. One can see this by realizing that in the discrete case,
in essence each point in $\mathcal{S}$ is the source of a microscopic mode in
$Z(\beta)$, and that notions of adjacency in $\mathcal{S}$ can be interpreted
collectively in terms of modes in $\mathcal{S}_\beta$ that contribute to
$Z(\beta)$.

\section{Design Analysis}
\label{sec:Analysis}
To better understand how the Pareto-Laplace filter encodes information about the
design problem we use $Z(\beta)$ to compute design information about moments of
$\mathcal{S}_\beta$, aspects of the transverse geometry, and consider questions
of solution robustness.

\subsection{Moments: Geometrical, Statistical, and Physical Aspects}
\label{sec:Moments}
Given the geometric interpretation of Eq.\ \eqref{eq:LTfamiliar} as the volume of a
solution space projection, the quantity
\begin{equation}
  \left<\mathcal{O}\right> = \frac{1}{Z(\beta)}
  \int_{\mathcal{O}_\text{min}}^{\infty}d\mathcal{O}
  e^{-\beta \mathcal{O}} \Omega_\perp(\mathcal{O}) \mathcal{O}
  = -\frac{\partial \ln Z(\beta)}{\partial \beta}
  \; ,
  \label{eq:Centroid}
\end{equation}
can be interpreted as the geometric centroid of the transformed solution space,
called $\mathcal{S}_\beta$.  Eq.\ \eqref{eq:Centroid} is precisely the
statistical expectation for $\mathcal{O}$ via Eq.\ \eqref{eq:entropy}. This
indicates that $Z(\beta)$ plays the role of a moment-generating function for
$\mathcal{S}_\beta$.

Note that in geometric, statistical, and physical settings, moments provide key
information about the representative entities they describe. Geometrically, Eq.\
\eqref{eq:Centroid} describes a centroid, whereas statistically it represents an
expectation value for samples drawn uniformly on $\mathcal{S}_\beta$.
Physically, Eq.\ \eqref{eq:Centroid} represents a thermal expectation value.

This intuition generalizes to cases with multiple objective functions. For
multi-objective cases
\begin{equation}
  \left<\mathcal{O}_i\right> = -\frac{\partial \ln Z}{\partial \beta_i}
  \label{eq:CentroidMulti}
\end{equation}
remains a geometric centroid, an expectation value, and a thermal average for
$O_i$. Likewise, higher-order moments
\begin{equation}
  \sqrt{\left<\mathcal{O}_i^2\right>-\left<\mathcal{O}_i\right>^2}
  = \sqrt{\frac{\partial^2 \ln Z}{\partial \beta_i^2}}
  \label{eq:SecondMomentMulti}
\end{equation}
represent characteristic sizes, as well as being related to statistical
variances, and thermal susceptibilities. Additionally, because the multiple
objectives supply added dimensions, it is possible to compute additional moments
\begin{equation}
\left<\mathcal{O}_i\mathcal{O}_j\right>-\left<\mathcal{O}_i\right>\left<\mathcal{O}_j\right>
  = \frac{\partial^2 \ln Z}{\partial \beta_i\partial \beta_j}
  \label{eq:SecondGeneral}
\end{equation}
that also encode aspects of the geometry of the solution space.

In mechanics, moments of the form Eq.\ \eqref{eq:SecondGeneral} relate to
angular motion, hence if the geometry of the solution space has the form of a spherical top this implies a strong symmetry of the geometry of the projected
$\mathcal{S}$ for given values of $\beta_i$. Statistically, that would imply a
lack of statistical correlation among the deviations of solutions from average
objective values. Moments that deviate from sphericity indicate hierarchical
relationships among objectives (i.e.\ that some have more variability than
others), or that one or more objectives are acting in concert or opposition to
one another.  Statistically, these would manifest in terms of
variance/covariance or correlation functions that follow taking the
point of view that Eq.\ \eqref{eq:SecondGeneral} defines a covariance matrix.

However, the physical representation of Eq.\ \eqref{eq:LTmulti} also indicates
that moments of the form Eq.\ \eqref{eq:SecondGeneral} can be thought of as
treating the transformed solution space as a piece of physical material and
determining the physical deformation of it to the application of anisotropic
pressure. Depending on the form of Eq.\ \eqref{eq:SecondGeneral} the solution
space could behave like an isotropic fluid (highest symmetry) or an anisotropic,
shear-supporting solid (least symmetry).

Each of these perspectives provides a window on questions of sensitivity, either
in terms of the ``squishiness'' of the solution space, it's geometric
dispersion, or its statistical covariance, that could provide useful insight
depending on the fluency of the practitioner and the problem in question.
The diversity of perspectives and the problems to which they could be applied is
too large to provide a representative survey here. Some of these perspectives
have been treated in special cases of this approach applied to materials design
and naval architecture, e.g.\ Refs.\ \cite{digitalalchemy,robustdesign}.

\subsection{Transverse Geometry}
The foregoing description of the solution space focused on geometric,
statistical, and physical understanding of the space as projected on an
axis that corresponds with the design objective $\mathcal{O}$. However, although
the quality of a design solution is captured in terms of $\mathcal{O}$, the form
of form of the solution is mostly specified in the transverse directions in
$\mathcal{S}$. Hence, knowledge of the geometric form of $\mathcal{S}$ in the
transverse space is crucial to questions about the structure of a design and
its realization. 

\subsubsection{Coarse Graining}
Understanding the geometry of the transverse space is most useful and
challenging in the case where the dimensionality of $\mathcal{S}$ is large. In
those situations it is useful \cite{robustdesign} to identify some overall
characteristics of putative design solutions (which could be a composite of basic
design elements) and to examine projections of the solution space geometry onto
those coordinates.

To enact this, one can start with $Z(\beta)$ in the form Eq.\
\eqref{eq:ZintIT} introduce a foliation of $\mathcal{S}$ in terms of both
$\mathcal{O}$ and some design characteristic $\mathcal{C}$. This yields
\begin{equation}
  Z(\beta) =
  \int d\mathcal{C}
  \int d\mathcal{O}
  e^{-\beta\mathcal{O}}
  \Omega_\perp(\mathcal{O},\mathcal{C}) \; ,
  \label{eq:Zoc}
\end{equation}
where $\Omega_\perp(\mathcal{O},\mathcal{C})$ is the volume of the slice in
$\mathcal{S}$ that is transverse to both $\mathcal{O}$ and $\mathcal{C}$.

In physics terminology, Eq.\ \eqref{eq:Zoc} implements a version of
coarse-graining because it effectively ``lumps together'' a set of states by
providing a description that does not explicitly depend on some of the
properties of the states. In conventional physical settings, this approach
is extremely powerful (see, e.g., Ref.\ \cite{goldenfeld}).

To enact this approach, it is convenient to work in terms of the so-called
Landau free energy, $F(\beta,\mathcal{C})$. The Landau free energy is given by
\begin{equation}
  e^{-\beta F(\beta,\mathcal{C})}
  =
  \int d\mathcal{O}
  e^{-\beta\mathcal{O}}
  \Omega_\perp(\mathcal{O},\mathcal{C}) \; ,
  \label{eq:LFEdef}
\end{equation}
so that
\begin{equation}
  Z(\beta) = \int d\mathcal{C}
  e^{-\beta F(\beta,\mathcal{C})} \; .
  \label{eq:ZLandau}
\end{equation}
In this form, the Landau free energy encodes an effective volume of the solution
space projected onto the direction of $\mathcal{C}$. Note that although we are
filtering the space in the $\mathcal{O}$ direction, Eq.\ \eqref{eq:LFEdef}
computes the effect of the $\mathcal{O}$ filter on a different characteristic of the
design, $\mathcal{C}$.

It can be particularly useful to examine the transverse geometry in cases
with multiple design objectives. For ease of illustration, we will consider two
objectives, but the generalization to an arbitrary number of objectives is
straightforward.

For two objectives we can redevelop Eq.\ \eqref{eq:Zoc} to get
\begin{equation}
  Z(\beta_1,\beta_2) =
  \int d\mathcal{C}
  \int d\mathcal{O}_1
  \int d\mathcal{O}_2
  e^{-\beta_1\mathcal{O}_1-\beta_2\mathcal{O}_2}
  \Omega_\perp(\mathcal{O}_1,\mathcal{O}_2,\mathcal{C}) \; .
  \label{eq:Zoc2}
\end{equation}
We can similarly consider the quantity
\begin{equation}
  e^{-\beta F(\beta_1,\beta_2,\mathcal{C})}
  =
  \int d\mathcal{O}_1
  \int d\mathcal{O}_2
  e^{-\beta_1\mathcal{O}_1-\beta_2\mathcal{O}_2}
  \Omega_\perp(\mathcal{O}_1,\mathcal{O}_2,\mathcal{C}) \; .
  \label{eq:LFEdef2}
\end{equation}

Eq.\ \eqref{eq:LFEdef2} plays an important role in algorithmic implementations
of this framework and for interpretation. In stochastic approaches to
optimization, particularly those that involve Markov Chains,
$-\nabla F(\beta_1,\beta_2,\mathcal{C})$ is an effective force in the solution
space for the sampling algorithm. In deterministic approaches,
$-\nabla F(\beta_1,\beta_2,\mathcal{C})$ encodes an average gradient across
$\mathcal{S}_\beta$ at fixed $\mathcal{C}$. Eq.\ \eqref{eq:LFEdef2} thus encodes
basic aspects of flows in algorithms. But because those flows are affected by
the competing pressures $\beta_{1,2}$, Eq.\ \eqref{eq:LFEdef2} is also important
for quantifying and interpreting trade-offs among objectives.

\subsubsection{Consequences of Objective Trade-Offs}
Since $Z$ retains its geometrical, statistical, and physical character in the
representation Eq.\ \eqref{eq:Zoc}, so Eq.\ \eqref{eq:LFEdef2} reduces the
effect of the competing pressures for objectives in terms the characteristic
$\mathcal{C}$. This means that Eq.\ \eqref{eq:LFEdef2} encodes the structure of
the design spaces as viewed from the perspective of the design characteristic
$\mathcal{C}$. Consequently, an analogous set of arguments can be employed
to quantify the geometry in terms of $\mathcal{C}$ as we showed for $\mathcal{O}$.
In particular $\left<\mathcal{C}\right>$ and
$\left<\mathcal{C}^2\right>-\left<\mathcal{C}\right>^2$ are geometrical and
statistical moments. Importantly, these moments are functions of $\beta_{1,2}$,
which means they express a relationship between some characteristic design
feature and the relationship among the design objectives. These sorts of
trade-offs have been evaluated in detail in, e.g., Ref.\ \cite{slo-go}.

\subsubsection{Effective Landscapes and Design Phases}
An additional useful insight from expressions of the form of Eq.\
\eqref{eq:LFEdef2} (and generalizations) is that they quantify the effective
form of the solution landscape as a function of one or more design
characteristics of interest. Because Eq.\ \eqref{eq:LFEdef2} is explicitly a
function of the competing design pressures, it encodes significant information
about the structure of the solution space both near global minima and away from
them. In physics applications quantities of this form are used to construct
phase diagrams that can distinguish distinct forms of the overall behaviour of
systems with macroscopically large numbers of degrees of freedom. The existence
of expressions of the form of Eq.\ \eqref{eq:LFEdef2} in design settings
indicates that similar approaches can be applied to identify analogues of phases in
design problems. Particular cases of this have already emerged in problems in
land-use planning \cite{Flashpoints} and in naval architecture.\cite{systemphys}

In presenting the framework although we could be agnostic about the precise form
of the design objectives and design characteristics, for the purposes of
interpretation we drew a conceptual distinction between them. However, from a
purely mathematical point of view, there is no reason why one could not develop
the framework above with the roles of the design objectives and an appropriately
quantified set of design characteristics reversed. In this way, the various
geometric, statistical, and physical lenses that can be used to focus on
relationships between objectives and characteristics could be used in both
directions. In colloquial terms, the approach we presented boils down to
``slicing'' and ``pinching'' the solution space in a systematic way. The only
thing that separates the application of this framework to optimization or
inverse optimization problems is ``angle'' of the slicing.

\subsection{Robustness}
\subsubsection{Near-Optimal Designs}
In many practical optimization problems, there is a limitation, e.g.\ finite
precision, that drives a mismatch between an as-designed optimal solution and
an as-realized real-world implementation. If any realization will inevitably miss
the ideal target, targets that are relatively more robust are ones for which
there exists a greater number of ways to have a near miss than ones in which
there are few ways to have a near miss. Hence basic questions about the
robustness of a putative solution depend on the near-optimal form of
$\mathcal{S}_\beta$.

To understand how this form of robustness is captured by the Pareto-Laplace
filter, consider a hypothetical situation in which there is a unique solution of
the optimization problem. If there is a unique solution at some
$\mathcal{O}_\text{min}$, then the volume of the solution space $\Omega_\perp$
will vanish for $\mathcal{O}<\mathcal{O}_\text{min}$. In that case, suppose that
for values of the compliance that fall just above $\mathcal{O}_\text{min}$ we can
approximate the phase space volume
\begin{equation}
  \Omega_\perp(\mathcal{O})=\gamma(\mathcal{O}-\mathcal{O}_\text{min})^{N_\text{IP}/\nu-1}
  \; ,
  \label{eq:OmegaNIP}
\end{equation}
where $\gamma$, $N_\text{IP}$, and $\nu$ are constants. What are they?

If the design problem has many variables (i.e.\ the dimensionality of
$\mathcal{S}\gg1$), it is not a priori clear that all of those variables will be
equally free near an optimal design. It could be that some number of those
variables are ``locked in'' and some of them are ``in play''. This asymmetry
between design, variables could originate from many sources: the unequal
distribution of the effects of constraints among the design variables or the
existence of inhomogeneity in the specification of the design objectives, among
others. Regardless of the source, the effective dimensionality of $\mathcal{S}$
near $\mathcal{O}_\text{min}$ could be less than the full dimensionality of
$\mathcal{S}$ for $\mathcal{O}\gg\mathcal{O}_\text{min}$. This can also be
thought of in physical terms: some of the degrees of freedom may ``condense''
for the design to enter the region of $\mathcal{S}$ in which $\mathcal{O}$ is
below some threshold, and the behaviour near $\mathcal{O}_\text{min}$ might
entail condensation of only the remaining degrees of freedom.

By this dimensional argument, the parametric growth of the near-optimal
$\Omega_\perp$ will be controlled by the number of degrees of freedom that are
``in-play'' near the minimum. Hence, $N_\text{IP}$ counts the effective number
of degrees of freedom that exhibit variation near optimality. The parameter
$\nu$ reflects how the objective function depends on variation in the
degrees of freedom. For example, for linear dependence $\nu=1$ whereas for
quadratic dependence $\nu=2$ (see Appendix \ref{ap:Scaling}). The coefficient
$\gamma$ is a geometric prefix that encodes the scaling of the growth.

Given this identification, it is possible to Eq.\
\eqref{eq:LTfamiliar} directly in the limit of large $\beta$ (small $T$)
\begin{equation}
  Z(\beta) \propto \frac{e^{-\beta \mathcal{O}_\text{min}}}{\beta^{N_\text{IP}/\nu}} \; ,
  \label{eq:Zapprox}
\end{equation}
where we have dropped negligible overall multiplicative constants. That means
that in the limit of large $\beta$ or, equivalently, low temperature $T$, we can
estimate Eq.\ \eqref{eq:Centroid} to be
\begin{equation}
  \left<\mathcal{O}\right> \approx \mathcal{O}_\text{min} + \frac{N_\text{IP}}{\nu} T \; .
  \label{eq:OlowT}
\end{equation}

Eq.\ \eqref{eq:OlowT} is an important relation. In particular, it may not be a
priori clear which degrees of freedom are in play near optimality. Indeed,
especially in non-linear problems, there may exist a complex set of correlations
among degrees of freedom that obscure the forms of variability that can exist.
Eq.\ \eqref{eq:OlowT} indicates that if there exists a means to evaluate
$\left<\mathcal{O}\right>$ near $T=0$ the slope of the curve determines the
number of degrees of freedom that are in play.

This can be viewed as a measure of robustness because it quantifies the space
available for ``near misses''. To see this, as the limit $\beta\to\infty$ or
$T\to0$ is approached, non-optimal solutions are effectively filtered out, while
near this limit, i.e., for sufficiently large $\beta$ or sufficiently small $T$
the geometry of the solution space leaves a clear imprint on the moments. Note
that similar reasoning can extend this beyond the small $T$ (large $\beta$)
limit. The analysis of such cases, which we give in Appendices
\ref{sec:GeoO} and \ref{ap:decrease}, shows that changes in the rate of growth
of $\Omega_\perp(\mathcal{O})$ yield identifiable effects on the behavior of
$\left<\mathcal{O}\right>$. Because these moments can be computed in algorithmic
implementations of the Pareto-Laplace filter, it is possible to obtain key
information about $\mathcal{O}_\perp$ even if it is not known directly.

\subsubsection{Modes}
One of the key conceptual outcomes of integral transforms in signal, control,
and other problems is the identification of characteristic modes of the system.
Here we derive analogous modes in the Pareto-Laplace framework.

Consider a situation in which there is a large density of solutions in
$\mathcal{S}$ at some value of the design objective $\mathcal{O}_*$ that fall in
the vicinity of some $x_\perp^{*}$ in the transverse space. It is instructive to
consider the contribution that these states would make to $Z(\beta)$, which we
will write as $Z_*(\beta,\mathcal{O}_*)$, which has the form
\begin{equation}
  Z_*(\beta,\mathcal{O}_*) = \int_{\mathcal{O}_\text{min}(\mathcal{C})}^{\infty}d\mathcal{O}
  e^{-\beta \mathcal{O}} \delta(\mathcal{O}-\mathcal{O}_*)
  \int_{V_*} d^{N-1}x_\perp \rho(x_\perp)\; ,
  \label{eq:ModeGeneral}
\end{equation}
where $\delta(\cdot)$ is the Dirac delta-function, $V_*$ is the volume of the
transverse space near $x_\perp^{*}$, and $\rho(x_\perp)$ is the density of
states on the region. If $\rho(x_\perp)$ is sufficiently well behaved over the
region that one can employ the mean value theorem, then Eq.\
\eqref{eq:ModeGeneral} reduces to
\begin{equation}
  Z_*(\beta,\mathcal{O}_*) = e^{-\beta \mathcal{O}_*} \bar{\rho} V_* \; ,
  \label{eq:Zstar}
\end{equation}
where $\bar{\rho}$ is the mean density of $\rho(x_\perp)$ over the region.

Eq.\ \eqref{eq:Zstar} indicates that $Z(\beta)$ encodes critical information
about the solution space, in general. To see this, note for fixed
$\mathcal{O}$, the regions in the solution space that contribute most to
$Z(\beta)$ are the ones that are most dense. Also note that for fixed density,
the regions of the space that fall off most slowly in $\beta$ are the ones for
which $\mathcal{O}$ is the smallest.

Another way of viewing this is by passing to the physical representation by
working in terms of $T$ in which case Eq.\ \eqref{eq:Zstar} becomes
\begin{equation}
  Z_*(T,\mathcal{O}_*) = e^{-\mathcal{O}_*/T} \bar{\rho} V_* \; .
  \label{eq:ZstarT}
\end{equation}
We can compare the relative contribution of two regions of identical density and
volume at two different values of the objective 
$\mathcal{O}_*^{(1)}$ and
$\mathcal{O}_*^{(2)}$ at fixed temperature as
\begin{equation}
  \frac{Z_*(T,\mathcal{O}_*^{(1)})}{Z_*(T,\mathcal{O}_*^{(2)})}
  =
  e^{(\mathcal{O}_*^{(2)}-\mathcal{O}_*^{(1)})/T}
  \; ,
  \label{eq:ZstarCompare}
\end{equation}
Note that Eq.\ \eqref{eq:ZstarCompare} is greater than one if
$\mathcal{O}_*^{(1)}<\mathcal{O}_*^{(2)}$. In other words, at any temperature,
if two regions of $\mathcal{S}$ have the same volume and density, the one that
contributes more to $Z(\beta)$ will be the one with lower $\mathcal{O}$.

Note that this property works in both directions. In particular, if there exists
a process for generating samples of states in $\mathcal{S}$ that satisfy the
statistical properties of Eq.\ \eqref{eq:LTfamiliar}, then, all else being
equal, the samples will more frequently display features that occur most
frequently in designs that minimize the objective. In this sense Eq.\
\eqref{eq:LTfamiliar} provides a window into the most important aspects of the
design space without needing to know the optimal design.

\subsubsection{Condensation}
It is also
instructive 
to consider the case in which certain regions of the solution space
are densely represented in the part of $\mathcal{S}$ for a band of
$\mathcal{O}$.

The first case we will consider is $\mathcal{O}_\text{min}\le \mathcal{O}\le
\mathcal{O}_\text{min}+\Delta\mathcal{O}$. For illustrative purposes, for the
moment, let's consider constant volume in $x_\perp$ over this band of
$\mathcal{O}$, i.e.\ $\Omega_\perp(\mathcal{O})=\Omega_\perp$. We can compute
the contribution of such a region to $Z(\beta)$, which will be given by
\begin{equation}
  Z(\beta,\Delta\mathcal{O})
  = \int_{\mathcal{O}_\text{min}}^{\mathcal{O}_\text{min}+\Delta\mathcal{O}}
  d\mathcal{O} \, e^{-\beta\mathcal{O}} \Omega_\perp \; .
  \label{eq:DeltaOmin}
\end{equation}
One could compute this integral in closed form, but it is more useful not to.
Instead, one can use the mean value theorem which dictates that there exists
some $\mathcal{O}_*$ that satisfies $\mathcal{O}_\text{min}\le \mathcal{O}_*\le
\mathcal{O}_\text{min}+\Delta\mathcal{O}$ such that
\begin{equation}
  Z(\beta,\Delta\mathcal{O})
  = \Delta\mathcal{O}
  \Omega_\perp
  e^{-\beta\mathcal{O}_*}
  \; .
  \label{eq:DeltaOstar}
\end{equation}
If $\Omega_\perp$ is not constant, but satisfies some meaningful notions of
continuity, then the same relationship holds, but with $\Omega_\perp$ replaced
by some appropriate notion of a mean.

How should one interpret this?

Suppose, for example, there are certain aspects of the design that characterize
all of the possible design choices in which the objective is within
$\Delta\mathcal{O}$ of the minimum. That set of design choices will materialize
as contributions to $Z(\beta)$ with a strength that is proportional both to the
range $\Delta\mathcal{O}$ for which they are common, and the volume
$\Omega_\perp$ of the transverse space the occupy, with a scaling factor that is
determined by the objective function and $\beta$. Note that because of the
exponential dependence of this scaling factor on $\beta$,
$\mathcal{O}_*\to\mathcal{O}_\text{min}$ as $\beta\to\infty$.

Now suppose there is a second region of $\mathcal{S}$ that is localized in a
region of the transverse space away from $\mathcal{O}_\text{min}$, e.g.,
$\mathcal{O}_\text{bad}\le \mathcal{O}\le
\mathcal{O}_\text{bad}+\Delta\mathcal{O}$. To facilitate comparison, we will
take $\Omega_\perp$ and $\Delta\mathcal{O}$ to be the same as before and
$\mathcal{O}_\text{bad}-\mathcal{O}_\text{min}\gg\Delta\mathcal{O}$. Integrating
to find the contribution to $Z(\beta)$ would yield the same expression as Eq.\
\eqref{eq:DeltaOstar} except that now
$\mathcal{O}_\text{bad}\le \mathcal{O}_*\le
\mathcal{O}_\text{bad}+\Delta\mathcal{O}$. The relative contribution of the two
regions to $Z(\beta)$ is differs exponentially the value of $\mathcal{O}_*$. In
particular, the ratio of the near minimal contribution to the non-minimal
contribution will be 
$\exp(-\beta(\mathcal{O}_\text{bad}-\mathcal{O}_\text{min}))
=\exp(-(\mathcal{O}_\text{bad}-\mathcal{O}_\text{min})/T)$.

Overall, suppose one is ignorant not only of $\mathcal{O}_\text{min}$ but also
the value of $x_\perp$ there, but still able to construct some way of sampling
the $\mathcal{S}_\beta$ according to Eq.\ \eqref{eq:LTfamiliar}. Regardless of whether
this sampling is generated geometrically, statistically, or physically, the
predominance of design features that characterize optimal solutions versus those
that are that occupy a similar (untransformed) volume of $\mathcal{S}$ will be
discernible for temperatures $T\lesssim
(\mathcal{O}_\text{bad}-\mathcal{O}_\text{min})$. In more colloquial terms, 
common elements that distinguish ``bad'' solutions from ``good'' ones (as scored
by $\mathcal{O}$) are culled from the solution space at higher $T$.

For example, if one was to generate samples on $\mathcal{S}_\beta$ by Markov-chain
Monte Carlo, the principle of detailed balance would indicate that if the
sampling was ergodic, key features of the design would begin to condense at
higher $T$. Similar effects would exist in other sampling methods. We will not
explore this at further length here, but a detailed account of this plays out in
practice in the context of structural design can be found in Ref.\
\cite{HyperoptMorphogenesis}.

\section{Illustrative Example: Linear Programming}
Although we are primarily interested in problems that don't admit closed-form
evaluation of the Pareto-Laplace filter, examples that do admit closed-form
evaluation are useful for illustrating key properties of $Z(\beta)$. This
section presents an example of a two-dimensional linear programming problem.
Appendix \ref{ap:GLP} considers general linear programming problems and Appendix
\ref{ap:Quad} considers quadratic programming.

For concreteness, consider the minimization problem:
\begin{equation*}
  \begin{split}
    \min \mathcal{O} &= -4x_1-3x_2+36\\
    \text{s.t.}\; & 3x_1+6x_2\le48\\
    & 4x_1+2x_2 \le 32\\
    &x_1+x_2\le 10\\
    &x_1\ge 0\\
    &x_2\ge 0\\
  \end{split}
\end{equation*}
For ease of illustration, it is useful exploit the symmetry properties of the
framework (see Sec.\ \ref{sec:symmetry}) to make a coordinate transformation
\begin{equation*}
  \begin{split}
    x_1 = -\frac{3}{5}x_\perp-\frac{4}{25}\mathcal{O}+6
    x_2 = \frac{4}{5}x_\perp-\frac{3}{25}\mathcal{O}+4
  \end{split}
\end{equation*}
which puts the optimal solution at $(x_\perp,\mathcal{O})=(2,0)$. The feasible
region is a convex polygon with its remaining vertices at
$(-\tfrac{22}{5},4)$,
$(\tfrac{14}{5},2)$,
$(\tfrac{34}{5},12)$, and
$(\tfrac{2}{5},36)$. From this we can compute $\Omega_\perp(\mathcal{O})$ for
each region. We find
\begin{equation}
  \Omega_\perp = 
  \begin{cases}
  \frac{5}{2}\mathcal{O} & 0\le\mathcal{O}\le2\\
  \frac{3}{2}\mathcal{O}+2 & 2\le\mathcal{O}\le4\\
  \frac{1}{4}\mathcal{O}+7 & 4\le\mathcal{O}\le12\\
  15-\frac{5}{12}\mathcal{O} & 12\le\mathcal{O}\le36\\
  \end{cases}
  \; .
  \label{eq:OmegaLP}
\end{equation}
We can then compute $Z(\beta)$ according to Eq.\ \eqref{eq:LTfamiliar} to get
\begin{equation}
  Z(\beta) =
  \frac{5}{2\beta^2}
  -\frac{1}{\beta^2}e^{-2\beta}
  -\frac{5}{4\beta^2}e^{-4\beta}
  -\frac{2}{3\beta^2}e^{-12\beta}
  +\frac{5}{12\beta^2}e^{-36\beta} \; .
  \label{eq:ZexLP}
\end{equation}

It is important to make some remarks about the structure of $Z(\beta)$. First,
each term has a factor of $\beta^{-2}$ that traces to the linear dependence of
$\Omega_\perp$ on $\mathcal{O}$ in each of the regions. Second the numerical
coefficient is related to the change in the rate of linear dependence in each of
the regions. Finally, each term has an exponential dependence that is determined
by the value of the objective at each of the basic feasible solutions, i.e., by the
vertices of the feasible region.

This latter property, i.e., the existence of a decay mode corresponding to each
basic feasible solution is an important property of the transform. In
other settings similar features in Laplace transforms correspond to
characteristic modes that capture essential behaviours of a system. Here we see
that, for linear programming, an analogous property emerges via the vertices of
the feasible region.

It is instructive to compute the moments of $Z(\beta)$, and to do this it is
convenient to write
\begin{equation}
  -\log Z =
  2\log\beta
  -\log\left(
    \frac{5}{2}
  \right)
  -\log\left(
    1
    -\frac{2}{5}e^{-2\beta}
    -\frac{1}{2}e^{-4\beta}
    -\frac{4}{15}e^{-12\beta}
    +\frac{1}{6}e^{-36\beta}
  \right)
  \; .
  \label{eq:logZLP}
\end{equation}
We can then compute $\left<\mathcal{O}\right>$ as
\begin{equation}
  \left<\mathcal{O}\right>=\frac{2}{\beta}
  -\frac{
    \frac{4}{5}e^{-2\beta}
    +4e^{-4\beta}
    +\frac{48}{15}e^{-12\beta}
    -6e^{-36\beta}}
  {1
    -\frac{2}{5}e^{-2\beta}
    -\frac{1}{2}e^{-4\beta}
    -\frac{4}{15}e^{-12\beta}
    +\frac{1}{6}e^{-36\beta} }
    \; .
  \label{eq:OvalLP}
\end{equation}

Based on the geometric arguments above, we would expect that
$\left<\mathcal{O}\right>=0$ as $\beta\to\infty$, and Eq.\ \eqref{eq:OvalLP}
clearly satisfies this property.

Our analysis of an example problem in linear programming revealed a particular
form for $Z(\beta)$ in Eq.\ \eqref{eq:ZexLP} that was determined by the
geometric features of the feasible region. We arrived at this result with the
assistance of an affine transformation of the feasible region. Though this
affine transformation was useful for illustrative purposes, in linear
programming problems of interest one should not expect to be able to compute
such a transformation without already knowing the solution. However, it is
important to note that although the affine transformation we used in the example
problem was convenient, it was not necessary. As we showed in Sec.\
\ref{sec:symmetry}, the Pareto-Laplace filter defined in Eq.\
\eqref{eq:LTfamiliar} has well defined properties under coordinate
transformations. These transformation properties imply that any means that could
generate $Z(\beta)$ in any formulation of the problem will yield a sum of
exponentially decaying modes $\exp(-\beta\mathcal{O}_v)$ where $\mathcal{O}_v$
is the objective evaluated at each of the basic feasible solutions that
correspond with the vertices of the feasible region.

Several things are interesting to note here. In Appendix \ref{ap:GLP} we extend
our our analysis to $\mathcal{S}$ in which the dimensionality of $x_\perp$ is
larger, and we compute $Z(\beta)$ by integrating piecewise in $\mathcal{O}$ over
the polytope that defines the feasible region. Each piecewise region corresponds
to a location where a new constraint takes hold because it results in a change
in the form of $\Omega_\perp(\mathcal{O})$. This break in the integration
induces a factor of $\exp(-\beta\mathcal{O})$ at a vertex (or in some special
cases, an edge) of the polytope. This means that (i) geometric features of the
solution space imprint themselves on the form of $Z(\beta)$ in a discernible
way, and that (ii) the analogue of long-lived transients in time-domain problems
are near optimal features in optimization problems.

\section{Discussion}
\begin{figure}
  \includegraphics[width=\textwidth]{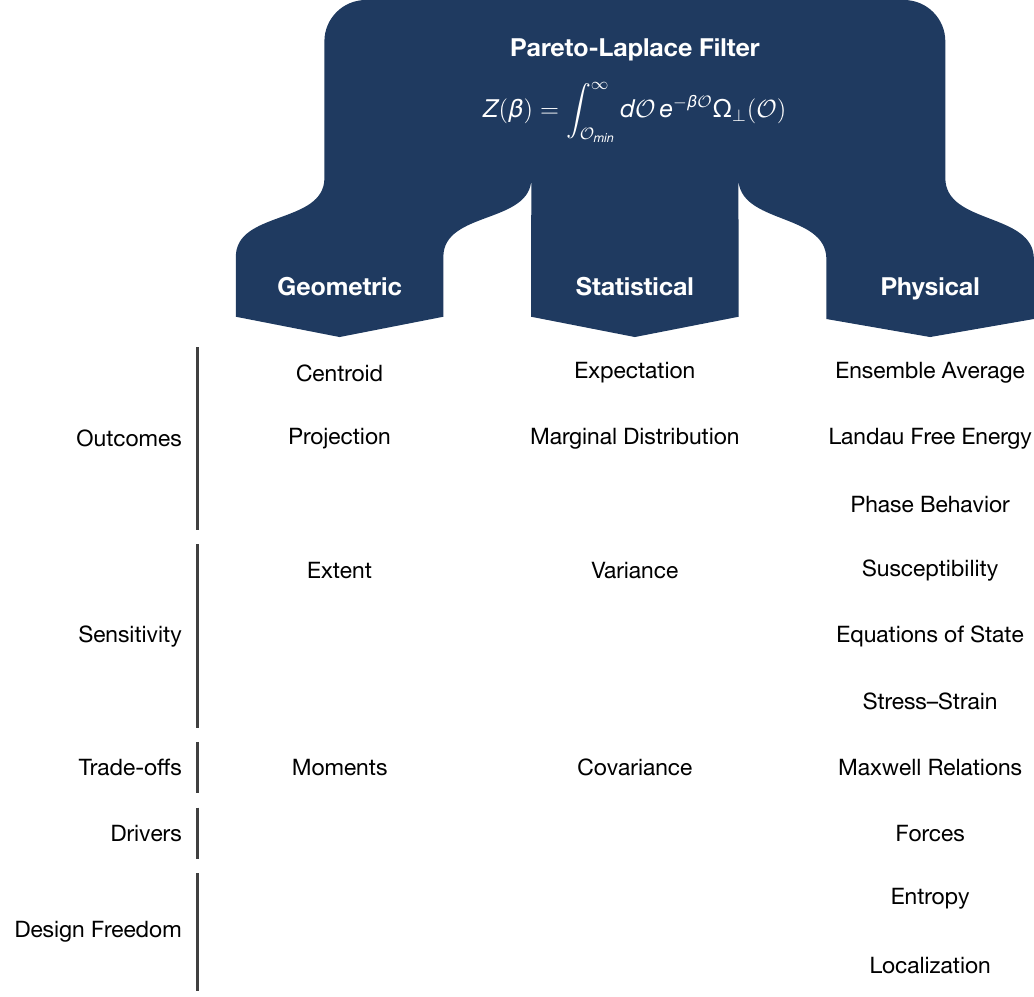}
  \caption{
    The Pareto-Laplace filtering framework can be interpreted geometrically,
    statistically, and physically. This framework provides the means to
    operationalize multiple different forms of design investigation in terms of
    one or more of the three perspectives.
  }
  \label{fig:Trichotomy}
\end{figure}

Prompted by the challenge of relating the structure of an optimization problem
to the structure of its solution, we constructed a Pareto-Laplace integral
transformation framework for design problems. We showed that the Pareto-Laplace
framework can be viewed from geometric, statistical, and physical perspectives
(Fig.\ \ref{fig:Trichotomy} illustrates this schematically). This multiplicity of
perspectives opens several windows on the relationship between problem- and
solution structure in optimization. We computed closed-form, explicit results in
some example cases, and showed how to construct a general formulation of the
approach for problems with an arbitrary number of objectives and constraints. We
also related our framework to other known approaches, some of which can be
understood as special cases.

The primary goal of the description we presented here was to establish some
basic properties of the framework. Because of the generality of the framework, it
is not possible to be exhaustive in describing all of the problems it can be applied
to, nor the forms of analysis that could leveraged. We concentrated our analysis
of the Pareto-Laplace framework by elucidating some of its key aspects using its
physical representation. We connected this framework to both geometry and
information theory, but on both fronts, we only invoked relatively primitive
tools. More sophisticated tools could be invoked to develop deeper geometric or
statistical connections and understanding. Moreover, because the framework is
a Laplace transform and because many disciplines have well-established
methodologies for handling and interpreting such transforms, practitioners may
find other fruitful ways of approaching this framework beyond the
physics-centric presentation we gave here.

The multiplicity of settings in which this framework can be invoked gives rise
to a corresponding multiplicity of means to implement it. We pointed to existing
implementations of this framework that leverage Monte
Carlo,\cite{digitalalchemy,engent,Flashpoints,slo-go} molecular
dynamics,\cite{digitalalchemy,Alch-MD} and tensor network
techniques.\cite{aaa,nfl} However, this list is in no way
exhaustive. Moreover, note that the techniques that have already been used to
implement the Pareto-Laplace framework were adapted from techniques that were
developed to attack other problems, and we believe that the generality of the
framework will yield implementations adapted from other existing algorithms.

Finally, we identified a number of existing results in the literature that are
examples of the present framework, including examples from the design of
self-assembled materials, from naval architecture, from land-use planning, and
from structural design. This list is not exhaustive, and we expect to report
soon on applications to problem spaces beyond this set.

\begin{acknowledgments}
We acknowledge the support of the Natural Sciences and Engineering Research
Council of Canada (NSERC) grants RGPIN-2019-05655 and DGECR-2019-00469.
\end{acknowledgments}

\appendix
\section{Weighted Additive Multi-Objective Optimization}
\label{ap:weight}
Although a feature of the present framework is the ability to address
multi-objective problems while remaining agnostic about the relative relation
among the objectives, it is instructive to illustrate how address problems in
which the relationship among the objectives is fixed.

In situations where there is an overall objective that arises as a linear sum of
the objectives, one can write an overall objective
\begin{equation}
  \mathcal{O} = \sum_i P_i \mathcal{O}_i \; ,
  \label{eq:OmultiL}
\end{equation}
where the $P_i$ are a set of constants with units of
$[\mathcal{O}]/[\mathcal{O}_i]$.

\subsection{Weight Parameterization and Competing Pressures}
\label{ap:CompetingP}
One can compute $Z(\beta)$ from Eq.\ \eqref{eq:OmultiL} by introducing
a parameter $\beta$ and employing Eq.\ \eqref{eq:LTfamiliar}. Note that, in
general, this recipe would involve $N_o+1$ parameters for $N_o$ objectives. The
additional parameter is essentially a free parameter that corresponds to an
arbitrary choice of units for $\mathcal{O}$ (the units of $\beta$ should be the
inverse of the units of $\mathcal{O}$). One could, for example, fix the units
of $\mathcal{O}$ to match those of one of the objectives, $\mathcal{O}_j$, set
the corresponding $P_j=1$, and the remaining $P_j$ would function akin to
``currency'' conversion factors between the objectives.

We note that factors, which are conventionally referred to as ``weights'' in
optimization, do not actually have the mathematical form of weight either in the
present framework or in optimization. In particular, whereas physical weight is an
extensive parameter, i.e., a parameters that scale with system size,
optimization ``weights'' are intensive. From the point of view of the present
framework, ``weights'' actually take the form of generalized pressures or
chemical potentials in the language of physics. In particular $P_i$ are the
mathematical analogues of pressure that quantify a form of competition between
objectives. This mathematical form aligns nicely with the notion of ``competing
pressures'' that is invoked in vernacular descriptions of multi-objective
problems. Further, note that in these cases
$\Omega_\perp=\Omega_\perp(\mathcal{O},P_i)$. The $P_i$ dependence arises
because the geometry of the Pareto slicing depends on functional form of
$\Omega$, which in turn depends on the choice of the competing pressures.

\subsection{Beyond Additive Multi-Objective Problems}
\label{ap:BeyondAdd}
What if the overall objective is not a linear combination of the component
objectives? Suppose the overall objective involved a product of objectives. In
these cases one can introduce a $\beta$ with units that are the inverse of the
units of the product, and then proceed to compute Eq.\ \eqref{eq:LTfamiliar}.

\section{Design Space Geometry Effects on Objective Function Moments}
\label{sec:GeoO}
Although the limit $\beta\to\infty$ or $T\to0$ filters out all
non-optimal solutions, near the strict limit, i.e., for sufficiently
large $\beta$ or sufficiently small $T$ the geometry of the solution space
leaves a clear imprint on the moments.

One situation is to understand the temperatures at which sub-leading corrections
are important in $\left<\mathcal{O}\right>(T)$. To this end, consider the case where there
is a single minimum at $\mathcal{O}_\text{min}$ and the volume in the vicinity is
\begin{equation}
  \Omega_\perp(\mathcal{O}) = \gamma
  (\mathcal{O}-\mathcal{O}_\text{min})^{N_\text{IP}/\nu-1}
  \left(1+\left(\frac{\mathcal{O}-\mathcal{O}_\text{min}}{T_*}\right)^{\Delta N}\right)
  \; ,
  \label{eq:OmegaExpansion}
\end{equation}
where the leading order behaviour matches that of Eq.\ \eqref{eq:OmegaNIP} and
includes sub-leading order corrections scaled by $T_*$, which is a constant with
the same units as the objective, and $\Delta N$ is the change in scaling of
$\Omega$ as the temperature increases. A volume of states of this form could
imply that a distance approximately $T_*$ in the $\mathcal{O}$ direction in
$\mathcal{S}$ some additional set of $\Delta N$ degrees of freedom become
``active'', for example.

Given $\Omega_\perp$ in the form in Eq.\ \eqref{eq:OmegaExpansion}, we can compute
the Laplace transform Eq.\ \eqref{eq:LTfamiliar} which gives
\begin{equation}
  Z(T)=\
  \gamma \Gamma\left(\frac{N_\text{IP}}{\nu}\right)
  e^{-\mathcal{O}_\text{min}/T} T^{N_\text{IP}/\nu}
  \left(
    1+
    \frac{\Gamma\left(\frac{N_{IP}}{\nu}+\Delta N\right)}
    {\Gamma\left(\frac{N_{IP}}{\nu}\right)}
      \left(\frac{T}{T_*}\right)^{\Delta N}
    \right) \; .
  \label{eq:Zexpansion}
\end{equation}
One could then, for example, compute the effect of Eq.\
\eqref{eq:OmegaExpansion} on $\mathcal{O}$ and find
\begin{equation}
  \left<\mathcal{O}\right> \approx \mathcal{O}_\text{min} + \frac{N_\text{IP}}{\nu} T
  +
  \frac{\Delta N T}{
    \frac{\Gamma\left(\frac{N_{IP}}{\nu}\right)}
    {\Gamma\left(\frac{N_{IP}}{\nu}+\Delta N\right)}
    \left(\frac{T_*}{T}\right)^{\Delta N}+1
  }
  \; .
  \label{eq:OvalCorr}
\end{equation}

For $T\ll T_*$, Eq.\ \eqref{eq:OvalCorr} recovers the
expected form of Eq.\ \eqref{eq:OlowT} near $\mathcal{O}_\text{min}$, i.e.,
\begin{equation}
  \left<\mathcal{O}\right> \approx \mathcal{O}_\text{min} + \frac{N_\text{IP} T}{\nu}
  \; .
  \label{eq:OvalCorrLowT}
\end{equation}
For $T\gg T_*$ Eq.\ \eqref{eq:OvalCorr} recovers a similar form to Eq.\ \eqref{eq:OlowT}
but with $N_\text{IP}/\nu\to N_\text{IP}/\nu+\Delta N$ given by
\begin{equation}
  \left<\mathcal{O}\right> \approx \mathcal{O}_\text{min} +
  \left(\frac{N_\text{IP}}{\nu}+\Delta N\right) T
  \; .
  \label{eq:OvalCorrHighT}
\end{equation}
This implies a change in
the slope of $\left<\mathcal{O}\right>$ around $T_*$, with the slope serving as
an indicator of the active degrees of freedom on either side of this transition.

\section{Geodesics: Minimal Length Descent to Objective Minima}
\label{ap:Geodesic}
For the purpose of interpretation or algorithms, it is useful to consider geodesics
on the filtered space. Taking the metric as in Eq.\ \eqref{eq:Metric} we can
parameterize a curve via the objective $\mathcal{O}$ and write the length of a
path from $(\mathcal{O}_1,x_\perp^{(1)})$ to
$(\mathcal{O}_2,x_\perp^{(2)})$ as
\begin{equation}
  L = \int_{\mathcal{O}_1}^{\mathcal{O}_2}d\mathcal{O}
  \sqrt{1+e^{-2\beta\mathcal{O}/(N-1)}\left(\frac{d\vec{x}_\perp}{d\mathcal{O}}\right)^2}
  \; ,
  \label{eq:Length}
\end{equation}
where we will take that the induced metric in the transverse space is flat, and
where we assume that $\mathcal{O}_2 > \mathcal{O}_1$. In this case, we get that
\begin{equation}
  \frac{
    e^{-2\beta\mathcal{O}/(N-1)}
    \frac{d\vec{x}_\perp}{d\mathcal{O}}
  }{
    1+e^{-2\beta\mathcal{O}/(N-1)}
    \left(\frac{d\vec{x}_\perp}{d\mathcal{O}}\right)^2
  }
  =
  \vec{v}_\perp
  \label{eq:v0def}
\end{equation}
where $\vec{v}_\perp$ is a set of $N-1$ constants. We can use this to find
\begin{equation}
  \frac{d\vec{x}_\perp}{d\mathcal{O}}
  =
  \frac{
    e^{\beta\mathcal{O}/(N-1)}
    \vec{v}_\perp
  }{
    \sqrt{
    e^{-2\beta\mathcal{O}/(N-1)}
    -
    \vec{v}_\perp^2
  }
  } \; ,
  \label{eq:Velocity}
\end{equation}
where
\begin{equation}
  |\vec{v}_\perp| <
  e^{-\beta\mathcal{O}_2/(N-1)} \; .
  \label{eq:SpeedLimit}
\end{equation}
Given this restriction, we can re-parameterize $\vec{v}_\perp$ as
\begin{equation}
  \vec{v}_\perp = \sin\alpha \hat{e}_\perp
  e^{-\beta\mathcal{O}_2/(N-1)} \; ,
  \label{eq:vdef}
\end{equation}
where $\hat{e}_\perp$ is a unit vector in the transverse space, and $\alpha$ is
the angle of inclination for a straight line connecting the two end points in
the limit $\beta\to 0$ (i.e., in the untransformed space).  This gives
\begin{equation}
  \frac{d\vec{x}_\perp}{d\mathcal{O}}
  =
  \frac{
    \sin\alpha e^{-\beta(\mathcal{O}_2-\mathcal{O})/(N-1)}
  }{
    \sqrt{
      e^{-2\beta\mathcal{O}/(N-1)}
      -
      \sin^2\alpha e^{-2\beta\mathcal{O}_2/(N-1)}
    }
  }\hat{e}_\perp
  \; .
  \label{eq:Vel}
\end{equation}
It is useful to note that Eq.\ \eqref{eq:Vel} implies that
$\tfrac{d\vec{x}_\perp}{d\mathcal{O}}$ grows exponentially in $\mathcal{O}$.
That, in turn, means that a minimal length path between
an arbitrary point $(\mathcal{O}',\vec{x}_\perp')$ in the transformed space
and the minimum
$(\mathcal{O}_\text{min},\vec{x}_\perp^\text{min})$
will converge rapidly in the transverse directions as it descends in
$\mathcal{O}$.

We would like to integrate Eq.\ \eqref{eq:Vel}, so it is convenient to first rearrange
\begin{equation}
  \frac{d\vec{x}_\perp}{d\mathcal{O}}
  =
  \hat{e}_\perp
  e^{\beta\mathcal{O}_2/(N-1)}
  \frac{
    e^{-2\beta(\mathcal{O}_2-\mathcal{O})/(N-1)}
  \sin\alpha
  }{
    \sqrt{
    1- e^{-2\beta(\mathcal{O}_2-\mathcal{O})/(N-1)} \sin^2\alpha
    }
  }
  \label{eq:XprimeForm}
\end{equation}
If we define $\Delta x_\perp=(x_\perp^{(2)}-x_\perp^{(1)})\cdot
\hat{e}_\perp$, this gives
\begin{equation}
  \Delta x_\perp
  =
  e^{\beta\mathcal{O}_2/(N-1)}
  \int_{\mathcal{O}_1}^{\mathcal{O}_2}d\mathcal{O}
  \frac{
    e^{-2\beta\mathcal({O}_2-\mathcal{O})/(N-1)}
  \sin\alpha
  }{
    \sqrt{
    1- e^{-2\beta(\mathcal{O}_2-\mathcal{O})/(N-1)} \sin^2\alpha
    }
  }
  \; .
  \label{eq:DeltaXpInt}
\end{equation}
It is convenient to make a change of variables
\begin{equation}
  \sin\theta = \sin\alpha e^{-\beta(\mathcal{O}_2-\mathcal{O})/(N-1)}
  \; ,
  \label{eq:COV}
\end{equation}
which gives
\begin{equation}
  \Delta x_\perp
  =
  \frac{N-1}{\beta}
  \frac{e^{\beta\mathcal{O}_2/(N-1)}}{\sin\alpha}
  \int_{\sin^{-1}(\sin\alpha e^{-\beta\Delta\mathcal{O}/(N-1)})}^{\alpha}
  d\theta \sin\theta
  \; ,
  \label{eq:DXperpTHint}
\end{equation}
where $\Delta\mathcal{O}=\mathcal{O}_2-\mathcal{O}_1$. Eq.\
\eqref{eq:DXperpTHint} relates the transverse displacement with respect to the
change in the objective to an effective change in an arc length imposed by the
geometric filter. Integrating this gives
\begin{equation}
  \Delta x_\perp
  =
  \frac{N-1}{\beta}
  \frac{e^{\beta\mathcal{O}_2/(N-1)}}{\sin\alpha}
  \left(
    \sqrt{
    1-e^{-2\beta\Delta\mathcal{O}/(N-1)}
    \sin^2\alpha
    } - \cos\alpha
  \right)
  \; .
  \label{eq:DXperp}
\end{equation}
In essence Eq.\ \eqref{eq:DXperp} is akin to a ``line of sight'' correction to
the inclination angle $\alpha$ between $(\mathcal{O}_1,x_\perp^{(1)})$ and
$(\mathcal{O}_2,x_\perp^{(2)})$ that is imposed by the spatial curvature of the
filter in $\mathcal{S}_\beta$.

\section{Example: Manifestations of Multiple Minima}
\label{ap:multi}
This feature is clear in the linear programming example we showed above, but it
is useful to consider how similar behaviours could emerge in non-convex problems.
To give a sample illustration of this, we will consider a simple problem that
represents the existence of multiple minima.

\begin{figure}
\begin{center}
  \includegraphics{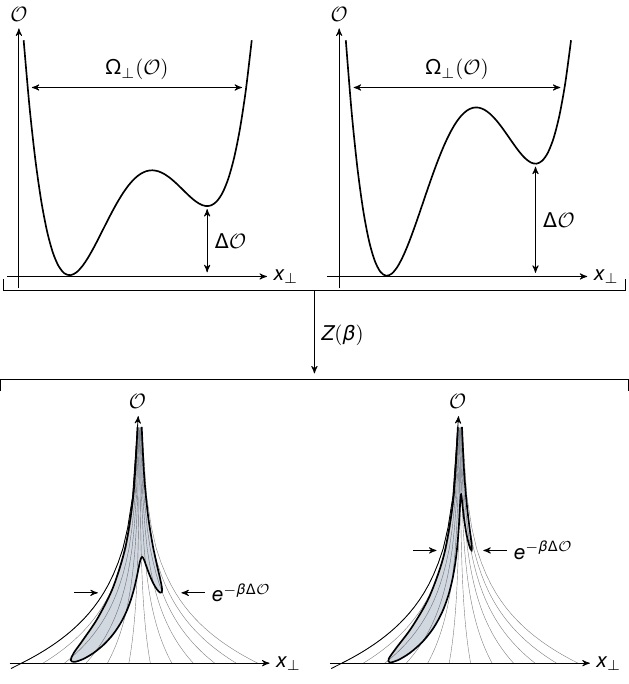}
\end{center}
  \caption{Schematic representation of Pareto-Laplace transform for a
    situation with multiple minima. In the pre-filtered picture (top panels),
    the sub-leading, local minimum is at a lower value of $\mathcal{O}$ in the
    scenario depicted in the left image compared to the scenario in the right
    image. In the post-filtered picture (lower panels) the filter more strongly
    ``pinches'' the region of the solution space near the local minimum with
    larger $\mathcal{O}$ (right) compared to the one with lower $\mathcal{O}$
    (left), illustrating the essence of the effect anticipated from
    Eq.\ \eqref{eq:Zmultiple}.
  }
  \label{fig:MultiMin}
\end{figure}

Suppose that there are two minima $\mathcal{O}_\text{min}^{(G,L)}$, where we are
assuming one is a local minimum and the other is a global minimum such that
$\mathcal{O}_\text{min}^{(G)} < \mathcal{O}_\text{min}^{(L)}$, and that there
are volumes associated with each that have the form
\begin{equation}
  \Omega_\perp(\mathcal{O}) =
  \gamma_G\theta(\mathcal{O}-\mathcal{O}_\text{min}^{(G)})
  (\mathcal{O}-\mathcal{O}_\text{min}^{(G)})^{N_\text{IP}^{(G)}/\nu-1}
  +
  \gamma_L\theta(\mathcal{O}-\mathcal{O}_\text{min}^{(L)})
  (\mathcal{O}-\mathcal{O}_\text{min}^{(L)})^{N_\text{IP}^{(L)}/\nu-1}
  \; ,
  \label{eq:OmegaMultiple}
\end{equation}
where $\gamma_{G,L}$ are constant coefficients, $\theta(\cdot)$ is the unit step
function which ensure the solution space volume contributions vanish below the
compliance minima, and $N_\text{IP}^{(G,L)}$ are the effective dimensions near
the respective minima. From this we can compute the Laplace transform
\begin{equation}
  Z(\beta) =
  \gamma_G \Gamma\left(\frac{N_\text{IP}^{(G)}}{\nu}\right) e^{-\beta \mathcal{O}_\text{min}^{(G)}} \beta^{-N_\text{IP}^{(G)}/\nu}
  +
  \gamma_L \Gamma\left(\frac{N_\text{IP}^{(L)}}{\nu}\right) e^{-\beta \mathcal{O}_\text{min}^{(L)}} \beta^{-N_\text{IP}^{(L)}/\nu}
  \; .
  \label{eq:Zmultiple}
\end{equation}
where $\Gamma(\cdot)$ is the gamma function. In terms of $T$ this takes the form
\begin{equation}
  Z(T) =
  \gamma_G \Gamma\left(\frac{N_\text{IP}^{(G)}}{\nu}\right) e^{-\beta \mathcal{O}_\text{min}^{(G)}} T^{N_\text{IP}^{(G)}/\nu}
  +
  \gamma_L \Gamma\left(\frac{N_\text{IP}^{(L)}}{\nu}\right) e^{-\beta \mathcal{O}_\text{min}^{(L)}} T^{N_\text{IP}^{(L)}/\nu}
  \; .
  \label{eq:ZmultipleT}
\end{equation}
Both of these forms illustrate that both minima leave their fingerprints as
modes in $Z(\beta)$ or $Z(T)$. If we rewrite Eq.\ \eqref{eq:ZmultipleT} as
\begin{equation}
\begin{split}
  Z(T) =&
  \gamma_G \Gamma\left(\frac{N_\text{IP}^{(G)}}{\nu}\right)
  e^{-\mathcal{O}_\text{min}^{(G)}/T} T^{N_\text{IP}^{(G)}/\nu}
  \\&
  \left(
    1
    +
    \frac{\gamma_L}{\gamma_G}
    \frac{\Gamma\left(\frac{N_\text{IP}^{(L)}}{\nu}\right)}{\Gamma\left(\frac{N_\text{IP}^{(G)}}{\nu}\right)}
    e^{-(\mathcal{O}_\text{min}^{(L)}-\mathcal{O}_\text{min}^{(G)})/T}
    T^{(N_\text{IP}^{(L)}-N_\text{IP}^{(G)})/\nu}
  \right)
  \; .
  \label{eq:ZmultipleTfact}
\end{split}
\end{equation}
we can see that the leading behaviour for small $T$ comes from the global
minimum, and that the effects of the local minimum on $Z(T)$ are suppressed by
the exponential factor
$\exp(-(\mathcal{O}_\text{min}^{(L)}-\mathcal{O}_\text{min}^{(G)})/T)$.
Fig.\ \ref{fig:MultiMin} gives an illustration of this phenomenon.

Note that it is possible to estimate the temperature at which $Z(T)$ goes from
being effectively characterized by a single minimum to being characterized by
two, which will occur for $T=T_\times$ such that
\begin{equation}
e^{-(\mathcal{O}_\text{min}^{(L)}-\mathcal{O}_\text{min}^{(G)})/T_\times}
  T_\times^{(N_\text{IP}^{(L)}-N_\text{IP}^{(G)})/\nu}
  \approx 
    \frac{\gamma_G}{\gamma_L}
    \frac{\Gamma\left(\frac{N_\text{IP}^{(G)}}{\nu}\right)}{\Gamma\left(\frac{N_\text{IP}^{(L)}}{\nu}\right)}
  \; .
  \label{eq:TwoAsOne}
\end{equation}
If the geometry near the two minima is similar, i.e., $N_{IP}^{(L)}\approx N_{IP}^{(G)}$
and $\gamma_L\approx\gamma_G$, then the crossover temperature is given by
\begin{equation}
  T_\times \approx \mathcal{O}_\text{min}^{(L)}-\mathcal{O}_\text{min}^{(G)} \; ,
  \label{eq:Tmulti}
\end{equation}
i.e.\ the objective difference of the two minima. This crossover temperature will
change if the growth of states away from the minima is much more rapid for one
minimum than the other. It is straightforward to extend the analysis above to
any number of minima.
\section{Example: General Linear Programming}
\label{ap:GLP}
In the text, we gave examples of the behaviour of Eq.\ \eqref{eq:LTfamiliar} for some
specific examples of linear programming. To further illustrate the behaviour of
$Z(\beta)$ in this section we consider a general linear programming problem in
$\mathbb{R}^n$.

\subsection{Pareto-Laplace Filter}
The general formulation of a linear programming problem with $m$ linear constraints is as Eq. (\ref{eq:LPGeneralFormulation}).
\begin{equation}
 \label{eq:LPGeneralFormulation}
 \begin{aligned}
\min_{x \in \mathbf{R}^n} \quad & \mathcal{O} = c^\top x + d_0\\
\textrm{s.t.} \quad & h^\top_j x + d_j \leq 0 \quad & j=1, \cdots, m \\
\end{aligned}
 \end{equation} 
The feasible region, $\mathcal{S}$, in linear programming, is a polytope made by the intersection of constraints where each is a hyperplane. The problem is to minimize the affine
function $\mathcal{O}(x)$ over this polytope. Given the fact that the objective function is linear, one can translate and rotate the coordinate such that the problem is reparametrized as Eq. (\ref{eq:RPLPGeneralFormulation}).
\begin{equation}
 \label{eq:RPLPGeneralFormulation}
 \begin{aligned}
\min_{x_\perp \in \mathbf{R}^{n-1}} \quad & \mathcal{O}\\
\textrm{s.t.} \quad & \Tilde{h}^\top_j x_\perp + h_0\mathcal{O} + \Tilde{d}_j \leq 0 \quad & j=1, \cdots, m \\
\end{aligned}
 \end{equation}

Without loss of generality, we assume the polytope $\mathcal{S}$ has no edge parallel to the $x_\perp$-plane and $N$ vertices. The vertices of the polytope $\mathcal{S}$ can be found by solving constraint equations. Defining $\Gamma_{\eta}$ as the $\mathcal{O}$-component of $\eta$\textsuperscript{th} vertex, the solution space $\Omega_\perp(\mathcal{O})$ is a piecewise function like Eq. (\ref{GeneralPiecewiseSolutionSpace}).
\begin{equation}
\label{GeneralPiecewiseSolutionSpace}
    \Omega_\perp(\mathcal{O}) = 
     \begin{cases}
       \sum_{i=0}^{n-1} a_i^{N-1}\mathcal{O}^i &\quad \Gamma_{N-1} \leq \mathcal{O} \leq \Gamma_N\\
        \quad\vdots &\quad \vdots\\
       \sum_{i=0}^{n-1} a_i^{1}\mathcal{O}^i &\quad \Gamma_{1} \leq \mathcal{O} \leq \Gamma_{2}\\
     \end{cases}
\end{equation}
Now, we can find the partition function of the problem as follows according to the definition given in Eq. (\ref{eq:LTfamiliar}).
\begin{equation}
\label{GeneralPartitionFunction}
\begin{split}
        Z(\beta) & = \sum_{\eta=1}^{N-1}\int_{\Gamma_{\eta}}^{\Gamma_{\eta+1}}d\mathcal{O}
  e^{-\beta \mathcal{O}} \left({\sum_{i=0}^{n-1} a_i^{\eta}\mathcal{O}^i}\right) \\
  & = \sum_{\eta=1}^{N-1}\mathcal{F}(\Gamma_{\eta},a^{\eta})e^{-\beta\Gamma_{\eta}} - \mathcal{F}(\Gamma_{\eta+1},a^{\eta})e^{-\beta\Gamma_{\eta+1}} \\
  & = \mathcal{F}(\Gamma_{1},a^{1})e^{-\beta\Gamma_{1}}\\
  & \quad + \sum_{\eta=2}^{N-1}\left[\mathcal{F}(\Gamma_{\eta},a^{\eta}) - \mathcal{F}(\Gamma_{\eta},a^{\eta-1})\right]e^{-\beta\Gamma_{\eta}}\\
  & \quad + \mathcal{F}(\Gamma_{N},a^{N-1})e^{-\beta\Gamma_{N}}
\end{split}
\end{equation}
where
\[
\mathcal{F}(\Gamma,a) = \sum_{j=0}^{n-1}\frac{1}{\beta^{j+1}}\frac{d^j}{d\Gamma^j}\left(\sum_{i=0}^{n-1}a_i\Gamma^i\right)
\]
Due to the continuity of $\Omega_\perp(\mathcal{O})$, we must have
\[
\sum_{i=0}^{n-1}a^{\eta}_i\Gamma_{\eta+1}^i = \sum_{i=0}^{n-1}a^{\eta+1}_i\Gamma_{\eta+1}^i
\]
This gives,
\[
\mathcal{F}(\Gamma_{\eta},a^{\eta}) - \mathcal{F}(\Gamma_{\eta},a^{\eta-1}) = \sum_{j=1}^{n-1} \frac{1}{\beta^{j+1}}\frac{d^j}{d\Gamma^j}\left( \sum_{i=0}^{n-1}(a_i^{\eta}-a_i^{\eta-1})\Gamma_{\eta}^i\right)
\]
Moreover, since $\Omega_\perp(\mathcal{O})$ between $\Gamma_{\eta} \leq \mathcal{O} \leq \Gamma_{\eta+1}$ is proportional to $(\mathcal{O}-\Gamma_\eta)^{n-1}$ for $\eta=1$ and $\eta=N-1$,
\[
\frac{d^j}{d\Gamma^j}\left(\sum_{i=0}^{n-1}a_{i}^{\eta}\Gamma_{\eta}^i\right) = 0
\]
for all $j \neq n-1$. Putting all of these together, we can simplify Eq. (\ref{GeneralPartitionFunction}) as Eq. (\ref{SimplifiedGeneralPartitionFunction}).

\begin{equation}
\label{SimplifiedGeneralPartitionFunction}
    Z(\beta) = \sum_{\eta=1}^N\zeta_\eta e^{-\beta\Gamma_\eta}
\end{equation}
where
\[
\begin{split}
    \zeta_1 &= \frac{(n-1)!}{\beta^n} a_{n-1}^1\\
    \zeta_\eta &= \sum_{j=1}^{n-1} \frac{1}{\beta^{j+1}} \frac{d^j}{d\Gamma^j} \left(\sum_{i=0}^{n-1} (a_i^{\eta} - a_i^{\eta-1}) \Gamma_{\eta}^i\right); \quad \quad \quad 2\leq \eta \leq N-1\\
    \zeta_N &= - \frac{(n-1)!}{\beta^n} a_{n-1}^{N-1}\\
\end{split}
\]
\subsection{Invariance}
According to the definition of $\zeta_\eta$, the partition function, $Z$ has no 0\textsuperscript{th} order coefficients, $a^{\eta}_0$, for any $\eta$. This means any translation of the coordinate along $x_{\perp}$ does not change the partition function. Moreover, since only the difference of the 1\textsuperscript{st} order coefficients, $a^{\eta}_1 - a^{\eta-1}_1$, exists, any rotation of the coordinate around $\mathcal{O}$-axis does not change the partition function. 
\[
x_\perp \rightarrow R_{\mathcal{O}}x_{\perp}+\Delta x_\perp \quad \quad \quad \Rightarrow \quad \quad \quad Z(\beta) \rightarrow Z(\beta)
\]
Although the translation of the coordinate by $\Delta\mathcal{O}$ along $\mathcal{O}$-axis does not change the geometry of $\mathcal{S}$, i.e.  $\zeta_\eta$ will be the same for all $\eta$, it changes the $\Gamma_\eta$ to $\Gamma_\eta-\Delta\mathcal{O}$.
\[
O \rightarrow {\mathcal{O}}+\Delta\mathcal{O}  \quad \quad \quad \Rightarrow \quad \quad \quad Z \rightarrow e^{\beta\Delta\mathcal{O}}Z(\beta)
\]
\subsection{Numerical Example: Three Dimensions}
For concreteness, we give a specific example in three dimensions. Consider the following problem,
\begin{equation*}
  \begin{split}
    \min \mathcal{O} &= 3x_1+4x_2-12x_3\\
    \text{s.t.} \quad & 5x_1+7x_2+7x_3\le 19\\
    & x_1 - x_3 \ge 0\\
    & x_2 - x_3 \ge 0\\
    & x_3 \ge 0 \, .\\
  \end{split}
\end{equation*}
For the matter of illustration, it is useful to make a coordinate transformation
such that $\mathcal{O}=\mathcal{O}(x_3)$, and $x_* = (0,0,0)$, where $x_*$ is
the global optimal point. To do this, we should rotate the coordinate by $\theta
= \arccos\left(
\frac{\overrightarrow{n}\cdot\overrightarrow{x_3}}{|\overrightarrow{n}||\overrightarrow{x_3}|}
\right)$ around vector $\overrightarrow{r} =
\overrightarrow{n}\times\overrightarrow{x_3}$, where $\overrightarrow{n}$ is the
normal vector of plane $\mathcal{O}$. This can be effected by the transformation:
\[
\begin{split}
    x &= \frac{4 x_1}{13}-\frac{12 x_2}{13}+\frac{3 x_3}{13}+1 \, ,\\
    y &= -\frac{12 x_1}{13}-\frac{3 x_2}{13}+\frac{4 x_3}{13}+1 \, ,\\
    z &= -\frac{3 x_1}{13}-\frac{4 x_2}{13}-\frac{12 x_3}{13}+1 \, .
\end{split}
\]
The transformed objective function is $\mathcal{O} = 13z-5$. Thus, we should set $z=\frac{1}{13}(\mathcal{O}+5)$. Then the optimization problem becomes 
\begin{equation*}
  \begin{split}
    \min & \quad \mathcal{O}\\
    \text{s.t.} & \quad \frac{1}{13} \left(-\frac{41}{13} (\mathcal{O}+5)-85 x-109 y\right)\le 0\\
    & \quad \frac{1}{13} \left(\frac{15 (\mathcal{O}+5)}{13}+7 x-8 y\right) \ge 0\\
    & \quad \frac{1}{13} \left(\frac{16 (\mathcal{O}+5)}{13}-9 x+y\right) \ge 0\\
    & \quad \frac{1}{169} (-12 \mathcal{O}-39 x-52 y+109) \ge 0 \, .\\
  \end{split}
\end{equation*}
Here $\vec{x}_\perp = (x,y)$. To find $\Omega_\perp$, we need to find the
equations of the lines that define the polygon that defines the boundary of the
feasible region in the $(x,y)$-plane. There are four constraints and
the intersection of each pair of them gives a line. One can find the following
line equations for this problem,
\begin{equation*}
    \begin{split}
        l_1(\mathcal{O}):&\left\{x = -\frac{151 \mathcal{O}}{1443}-\frac{755}{1443},y = \frac{76 \mathcal{O}}{1443}+\frac{380}{1443}\right\}\\
        l_2(\mathcal{O}):&\left\{x = \frac{131 \mathcal{O}}{1066}+\frac{655}{1066},y = -\frac{133 \mathcal{O}}{1066}-\frac{665}{1066}\right\}\\
        l_3(\mathcal{O}):& \left\{x = \frac{88 O}{13}-\frac{977}{13},y = \frac{760}{13}-\frac{69 O}{13}\right\}\\
        l_4(\mathcal{O}):& \left\{x = \frac{11 O}{65}+\frac{11}{13},y = \frac{19 O}{65}+\frac{19}{13}\right\}\\
        l_5(\mathcal{O}):& \left\{x = \frac{11}{13}-\frac{3 O}{13},y = \frac{19}{13}-\frac{3 O}{52}\right\}\\
        l_6(\mathcal{O}):& \left\{x = \frac{4 O}{39}+\frac{11}{13},y = \frac{19}{13}-\frac{4 O}{13}\right\}\\
    \end{split}
\end{equation*}
These lines intersect at the vertices of the feasible region for each
$\mathcal{O}$. For instance, the intersection of $l_1$ and $l_2$ gives the
$K:(0,0,-5)$. Thus, $\Gamma_1=-5$
\begin{equation*}
    \begin{split}
        \Gamma_1=&-5\\
        \Gamma_2=&0\\
        \Gamma_3=&\frac{76}{7}\\
        \Gamma_4=&\frac{57}{5} \, .\\
    \end{split}
\end{equation*}
Between $\Gamma_1$ and $\Gamma_2$, for each $\mathcal{O}$, $\Omega_\perp$ is the
area of a triangle bounded by $l_1$, $l_2$, and $l_4$. Defining
$\vec{e}_{ij}$ as the vector starts on $l_i$ and ends on $l_j$ lines for the same
$\mathcal{O}$, we can find the area of the triangle as
$\frac{1}{2}|\vec{e}_{12} \times \vec{e}_{14}|$. In the
same way, between $\Gamma_2$ and $\Gamma_3$, and between $\Gamma_3$ and
$\Gamma_4$, $\Omega_\perp$ is the area of a trapezoid, and a triangle, with area
of $\frac{1}{2}|\vec{e}_{16} \times \vec{e}_{25}|$, and
$\frac{1}{2}|\vec{e}_{26} \times \vec{e}_{23}|$,
respectively.
Finally, the $\Omega_\perp$ is as Eq. \ref{eq:Omega3DLP}.
\begin{equation}
  \Omega_\perp = 
  \begin{cases}
  \frac{4693 (\mathcal{O}+5)^2}{91020} & -5\le\mathcal{O}\le0\\
  -\frac{689 \mathcal{O}^2}{12136}+\frac{4693 \mathcal{O}}{9102}+\frac{23465}{18204} & 0\le\mathcal{O}\le\frac{76}{7}\\
  \frac{13}{492} (57-5 \mathcal{O})^2 & \frac{76}{7}\le\mathcal{O}\le\frac{57}{5}\\
  \end{cases}
  \; .
  \label{eq:Omega3DLP}
\end{equation}
Then using Eq.(\ref{SimplifiedGeneralPartitionFunction}), we have
\begin{equation}
    Z(\beta)= \frac{1}{\beta^3}\frac{4693}{45510}e^{5\beta} - \frac{1}{\beta^3}\frac{13}{60} + \frac{1}{\beta^3}\frac{637}{444}e^{-(76/7)\beta} - \frac{1}{\beta^3}\frac{325}{246}e^{-(57/5)\beta} \, .
\end{equation}
Note that in this form, $Z(\beta)$ has four modes with exponential dependence on
$\beta$ according to $\mathcal{O}$ evaluated at each basic feasible solution, i.e.\
each vertex of the polyhedron bounding the feasible region. The exponent of the polynomial coefficient, $\beta^{-3}$, corresponds to the dimensionality of the problem,
and there is a numerical factor determined by the geometry of the feasible region
between basic feasible solutions.

\section{Example: Quadratic Problems}
\label{ap:Quad}
\subsection{Simple Case}
Although evaluating Eq.\ \eqref{eq:LTfamiliar} for most nonlinear problems could
leverage numerical techniques such as Monte Carlo, molecular dynamics, or tensor
network methods, it may be instructive to consider a case in which it can be
computed exactly.

Consider a case where $\mathcal{S}=\{(x_1,x_2)|x_{1,2}\in\mathbb{R}\}$,
$\mathcal{O}=(x_1^2+x_2^2)$, and
\begin{equation}
    \Omega_\perp(\mathcal{O}) =
    \int_{-\infty}^{\infty} dx_1
    \int_{-\infty}^{\infty} dx_2 \delta(\mathcal{O}-(x_1^2+x_2^2)) \; .
\end{equation}
If we make the change of variables
\begin{equation}
  x_1 = r\cos\theta \qquad x_2=r\sin\theta
  \; ,
  \label{eq:QuadPolar}
\end{equation}
we have that the volume of the solution space transverse to the objective at
some particular $\mathcal{O}$ is
\begin{equation}
  \Omega_\perp(\mathcal{O}) =
  \int_0^{2\pi} d\theta\int_0^\infty dr r \delta(\mathcal{O}-r^2)
  \; .
  \label{eq:QuadOmegaInt}
\end{equation}
Using the delta-function identity
\begin{equation}
  \delta(\mathcal{O}-r^2) =
  \frac{\delta(r-\sqrt{\mathcal{O}})}{2 \sqrt{\mathcal{O}}}
    \; ,
  \label{eq:deltafnid}
\end{equation}
yields
\begin{equation}
  \Omega_\perp(\mathcal{O}) = \pi
  \; .
  \label{eq:QuadOmega}
\end{equation}
We get then that
\begin{equation}
  Z(\beta) = \frac{\pi}{\beta}
  \; .
  \label{eq:QuadZ}
\end{equation}
Note that this form gives $Z(\beta)$ as a single mode with the value of the
objective function taking the value of $0$ at the minimum, $\pi$ is clearly a
geometric coefficient that describes the growth of the solution space with the
objective, and $Z(\beta) \propto (\beta)^{-1}$ because volume of the transverse
space is constant.

\subsection{General Quadratic Programming}
Consider a more general quadratic program in $\mathbb{R}^N$ with an
objective function of the form
\begin{equation}
    \mathcal{O} = \frac{1}{2}x^T A x - b^T x \; ,
\end{equation}
where $A$ is a symmetric, positive definite matrix, and $b\in\mathbb{R}^N$.
In this case Eq.\ \eqref{eq:LTfamiliar} gives
\begin{equation}
    Z(\beta) = \int d^N x
    e^{-\beta(\frac{1}{2}x^T A x - b^T x)} \; .
\end{equation}

\subsubsection{Evaluating the Transform}
To evaluate this, we make a change of variables
\begin{equation}
    x = y+A^{-1}b \; ,
    \label{eq:Q-CoV}
\end{equation}
which gives $Z(\beta)$ as
\begin{equation}
    Z(\beta) =
    e^{-\frac{\beta}{2}b^T A^{-1}b}
    \int d^N y
    e^{-\frac{\beta}{2}y^T A y} \; .
\end{equation}
Standard formulae (see, e.g., Ref.\ \cite{zinnjustin}) then give
\begin{equation}
\label{eq:Zqdet}
    Z(\beta) =
    \left(\frac{2\pi}{\beta}\right)^{N/2}
    \mathrm{det}(A)^{-1/2}
    e^{-\frac{\beta}{2}b^T A^{-1}b}
    \; .
\end{equation}
If we suppose that there is a set of unit eigenvectors
\begin{equation}
    A \hat{\lambda}_i = \lambda_i \hat{\lambda}_i \; ,
\end{equation}
and we take
\begin{equation}
    b = \sum_i b_i \hat{\lambda}_i \; ,
\end{equation}
then Eq.\ \eqref{eq:Zqdet} takes the form
\begin{equation}
    Z(\beta) =
    \left(\frac{2\pi}{\beta}\right)^{N/2}
    \left(\prod_i \lambda_i^{-1/2}
    \right)
    e^{-\frac{\beta}{2}\sum_i \frac{b_i^2}{\lambda_i}}
    \; .
    \label{eq:Zquad}
\end{equation}
We can interpret this quantity as follows. Since the problem is
globally convex, there is a single mode, that corresponds to the
global minimum. The growth of solutions around the minimum is
determined by the dimensionality of the problem ($N$) and the spectrum
of $A$ (via $\lambda_i$). The weight of the mode is also determined by the
spectrum (again via $\lambda_i$), and by the location of the global minimum (via $b_i$).

\subsubsection{Near-Optimal Behaviour}
Given $Z(\beta)$ in Eq.\ \eqref{eq:Zquad} one can consider the near
optimal behaviour of the objective $\mathcal{O}$. We can deduce the
near optimal behaviour via Eq.\ \eqref{eq:Centroid}, which gives
\begin{equation}
    \left<\mathcal{O}\right>
    = \frac{1}{2} b^T A^{-1} b
    +\frac{1}{2}NT \; ,
\end{equation}
where we have used $T=1/\beta$.
Note that the temperature dependence is an example of what is referred
to in physics as the equipartition theorem (see, e.g., Ref.\ \cite{LLv5}).

\subsubsection{Transverse Geometry}
We evaluated Eqs.\ \eqref{eq:LTfamiliar} and \eqref{eq:Centroid} without the
need to decompose $\mathcal{S}$ to explicitly determine $\Omega_\perp$. However,
it may be instructive to show how to effect the decomposition. In particular, it is useful to understand the origin of the $\nu$ factor
in various scaling relations.

The transverse volume is given by
\begin{equation}
    \Omega_\perp(\mathcal{O}) =
    \int d^N x \, \delta\left(
    \mathcal{O} - \frac{1}{2}x^T A x + b^T x\right) \; .
\end{equation}
We first make the change of variables in Eq.\ \eqref{eq:Q-CoV} which gives
\begin{equation}
    \Omega_\perp(\mathcal{O}) =
    \int d^N y \, \delta\left(
    \mathcal{O} - \frac{1}{2}y^T A y\right) \; .
\end{equation}
We will assume that $A$ can be diagonalized by a similarity transformation
in $Q\in SO(N)$, and make another change of variables $u=Qy$, which gives
\begin{equation}
    \Omega_\perp(\mathcal{O}) =
    \int d^N u \, \delta\left(
    \mathcal{O} - \frac{1}{2}u^T \Lambda u\right) \; ,
\end{equation}
where $\Lambda$ is a diagonal matrix of the eigenvalues of $A$. We then make
yet another change of variables by defining $v\in \mathbb{R}^N$, such that
\begin{equation}
    v_i = \sqrt{\frac{\lambda_i}{2}} u_i \; ,
\end{equation}
which gives
\begin{equation}
    \Omega_\perp(\mathcal{O}) =
    \frac{2^{N/2}}{\mathrm{det}(A)^{1/2}}
    \int d^N v \, \delta\left(
    \mathcal{O} - v^2 \right) \; .
\end{equation}
We make a final change of variables to polar coordinates to get
\begin{equation}
    \Omega_\perp(\mathcal{O}) =
    \frac{2^{N/2}}{\mathrm{det}(A)^{1/2}}
    \int d\omega_{N-1}
    \int dr r^{N-1} \, \delta\left(
    \mathcal{O} - r^2 \right) \; ,
\end{equation}
where $r^2=v^2$, and $d\omega$ is the volume measure for the solid angle.
We can then apply the identity in Eq.\ \eqref{eq:deltafnid}, to get
\begin{equation}
    \Omega_\perp(\mathcal{O}) =
    \frac{2^{N/2}}{\mathrm{det}(A)^{1/2}}
    \int d\omega_{N-1}
    \int dr r^{N-1} \,
    \frac{ \delta(r - \sqrt{\mathcal{O}})}{2\sqrt{\mathcal{O}}} \; .
\end{equation}
Using the fact that
\begin{equation}
    \int d\omega_{N-1}
    =
    \frac{2\pi^{N/2}}{\Gamma\left(\frac{N}{2}\right)} \; ,
\end{equation}
we get that
\begin{equation}
    \Omega_\perp(\mathcal{O}) =
    \frac{(2\pi)^{N/2}}{\mathrm{det}(A)^{1/2}
    \Gamma\left(\frac{N}{2}\right)}
    \mathcal{O}^{\frac{N}{2}-1} \; .
\end{equation}
If one inserts this form into Eq.\ \eqref{eq:LTfamiliar} and notes that
\begin{equation}
    \mathcal{O}_\text{min} = \frac{1}{2}b^T A b
\end{equation}
then one recovers Eq.\ \eqref{eq:Zquad}.

Note that compared with the analysis of the general linear program in
Appendix \ref{ap:GLP}, the for the quadratic program here the difference
in the scaling $\nu$ arises from the difference in parametric dependence
of the objective function on the design variables $x$.

\section{Example: Non-Analytic Volume--Objective Problem}\label{ap:decrease}
We gave detailed analyses of cases in which the volume of Pareto-slices of
solution space depend analytically on the objective function. There may be cases
in which the Pareto-slice volume is a continuous function of the objective, but
it does not have continuous derivatives. Here, examine an example case in which
$\Omega_\perp(\mathcal{O})$ is defined piecewise.

Consider the case where some sector of $\mathcal{S}$ is ``in play'' near the
minimum, up to some saturation point, $\mathcal{O}_*$, in $\mathcal{O}$, after which a
different set of degrees of freedom comes into play. We represent this by
\begin{equation}
  \Omega_\perp(\mathcal{O}) =
  \begin{cases}
    \gamma_< (\mathcal{O}-\mathcal{O}_\text{min})^{N_</\nu_<-1} & \mathcal{O}<O_* \\
    \gamma_< (\mathcal{O}_*-\mathcal{O}_\text{min})^{N_</\nu_<-1}+
    \gamma_> (\mathcal{O}-O_*)^{N_>/\nu_>-1} & \mathcal{O}>O_* \\
  \end{cases}
  \; ,
  \label{eq:Opiece}
\end{equation}
where $\gamma_{<,>}$ are geometric coefficients, and $N_{<,>}$ are scaling
exponents giving the number of effective degrees of freedom for $\Omega_\perp$
on either side of $\mathcal{O}_*$, and $\nu_{<,>}$ index the growth in
$\mathcal{O}$ (e.g., linear, quadratic, etc.).

Integrating Eq.\ \eqref{eq:Opiece} gives
\begin{equation}
  \begin{split}
    Z(\beta) =&
    e^{-\beta\mathcal{O}_\text{min}}\beta^{N_</\nu_<}
    \gamma_<
    \left(\Gamma(N_</\nu_<)-\Gamma(N_</\nu_<,\beta(\mathcal{O}_*-\mathcal{O}_\text{min})\right)
      +\\
      & e^{-\beta\mathcal{O}_*}\beta^{N_>/\nu_>}
      \left(
        \gamma_>\Gamma(N_>/\nu_>)
        +\gamma_<\left[\beta(\mathcal{O}_*-\mathcal{O}_\text{min})\right]^{N_>/\nu_>}
      \right) \; .
    \end{split}
  \label{eq:Zpiece}
\end{equation}
where $\Gamma(\cdot,\cdot)$ is the incomplete Gamma function.

We can use this to find that for
$\beta(\mathcal{O}_*-\mathcal{O}_\text{min})\gg 1$
(equivalently $T\ll\mathcal{O}_*-\mathcal{O}_\text{min}$)
\begin{equation}
  \left<\mathcal{O}\right>\approx
\mathcal{O}_\text{min}+ \frac{N_<}{\nu_<\beta}
  = 
  \mathcal{O}_\text{min}+\frac{N_< T}{\nu_<}
  \; ,
  \label{eq:OTsPiece}
\end{equation}
whereas for
$\beta(\mathcal{O}_*-\mathcal{O}_\text{min})\ll 1$
(equivalently $T\gg\mathcal{O}_*-\mathcal{O}_\text{min}$)
\begin{equation}
  \left<\mathcal{O}\right>\approx
  \mathcal{O}_*+ \frac{N_>}{\nu_>\beta}
  =
  \mathcal{O}_*+ \frac{N_> T}{\nu_>}
  \; .
  \label{eq:OTlPiece}
\end{equation}

Note that for $T\ll\mathcal{O}_*-\mathcal{O}_\text{min}$ and
$T\gg\mathcal{O}_*-\mathcal{O}_\text{min}$,
$\mathcal{O}$ asymptotes to linear response in $T$. In both cases,
the slope is determined by the power law growth of $\Omega_\perp$. The
exponent in this power law growth is, in turn, determined by the number of
degrees of freedom that are ``in play'' (i.e., whose variation is subject to
filtering) at that level of $T$ (or $\beta$).

\section{Special Case: Simulated Annealing}
We gave explicit results in a set of problems that could be done
in closed form. However, the vast majority of problems of interest do not
admit closed-form solutions. We have applied the Pareto-Laplace framework
in several cases, which we have described elsewhere, e.g., Refs.\ \cite{Alch-MD,digitalalchemy,packingassembly,identitycrisis,systemphys,robustdesign,engent,
Flashpoints,slo-go}. However, the approach taken in those works differs
somewhat with some conventional optimization approaches, so it is
instructive to establish a more concrete connection with other approaches.

In this appendix we will describe the relationship of this approach to
simulated annealing. Simulated annealing is the optimization approach that
shares the closest kinship with the Pareto-Laplace framework. The great
utility of simulated annealing for a range of problems led the extension of the
original method described in Ref.\ \cite{kirkpatrick1983} in a large number of
ways. It is not possible to describe each of them here, so we will concentrate
on conventional simulated annealing, and we will leave the discussion of
extensions to other work.

In the language of the Pareto-Laplace framework, conventional simulated
annealing generates a random walk on $\mathcal{S}_\beta$, where $\beta$ is
``slowly'' (in some sense that is determined by the landscape of the problem)
increased from $\beta=0$ to $\beta\to\infty$. This protocol tends to converge
to the global minimum of the optimization problem, in the Pareto-Laplace
framework, because increasing $\beta$ effectively ``sucks the air'' out
of $\mathcal{S}_\beta$ for large $\mathcal{O}$. For large but finite
$\beta$ the remaining finite volume of $\mathcal{S}_\beta$ is concentrated
around $\mathcal{O}_\text{min}$.

The Pareto-Laplace framework adds several elements to simulated annealing. 
Some of these elements are: (1) The framework is agnostic about processes
that occur on $\mathcal{S}_\beta$. (2) The aim of the framework is to
characterize the entiretly of the structure of $\mathcal{S}_\beta$, and
its various parametric dependencies, rather than focusing on determining
optima. (3) The framework adds additional tools, e.g., modes, moments, etc.,
that aid in the interpretation of optimization results. (4) The framework
seeks to situate optimization problems in broader geometric, statistical,
and physical context for the purpose of opening up opportunities for the
use of tools from those domains of knowledge.

\section{Near-Optimal Scaling}
\label{ap:Scaling}
Here we derive the scaling of $\Omega_\perp$ in the vicinity of a minimum of
arbitrary index $\nu$. We assume that we interested in
$\mathcal{O}-\mathcal{O}_\text{min}\approx 0$ so that we can approximate
$\mathcal{S}_\beta$ locally as $\mathbb{R}^{N_\text{IP}}$ near the minimum,
which we take to be at $x_0$.

In this setting we have that
\begin{equation}
  \Omega_\perp(\mathcal{O}) =
  \int d^{N_\text{IP}} x \, \delta(\mathcal{O}-|x-x_0|^\nu) \; .
  \label{eq:OnuDef}
\end{equation}
Defining $y=x-x_0$ gives
\begin{equation}
  \Omega_\perp(\mathcal{O}) =
  \int d^{N_\text{IP}} y \, \delta(\mathcal{O}-|y|^\nu) \; .
  \label{eq:OnuY}
\end{equation}
We can then work in polar coordinates and take $r=|y|$, which gives
\begin{equation}
  \Omega_\perp(\mathcal{O}) =
  \int_0^{R} dr r^{N_\text{IP}-1} \, \delta(\mathcal{O}-r^\nu)
  \int_{S^{N_\text{IP}-1}} d\omega
    \; ,
  \label{eq:OnuR}
\end{equation}
where we assume $R>\mathcal{O}^{1/\nu}$.

Standard identities allow the evaluation of Eq.\ \eqref{eq:OnuR} as
\begin{equation}
  \Omega_\perp(\mathcal{O}) =
  \frac{2\pi^{N_\text{IP}/2}}{\nu\Gamma\left(\frac{N_\text{IP}}{2}\right)}
   \mathcal{O}^{\frac{N_\text{IP}}{\nu}-1}
   \; .
   \label{eq:OnuScale}
\end{equation}


\begin{thebibliography}{30}%
\makeatletter
\providecommand \@ifxundefined [1]{%
 \@ifx{#1\undefined}
}%
\providecommand \@ifnum [1]{%
 \ifnum #1\expandafter \@firstoftwo
 \else \expandafter \@secondoftwo
 \fi
}%
\providecommand \@ifx [1]{%
 \ifx #1\expandafter \@firstoftwo
 \else \expandafter \@secondoftwo
 \fi
}%
\providecommand \natexlab [1]{#1}%
\providecommand \enquote  [1]{``#1''}%
\providecommand \bibnamefont  [1]{#1}%
\providecommand \bibfnamefont [1]{#1}%
\providecommand \citenamefont [1]{#1}%
\providecommand \href@noop [0]{\@secondoftwo}%
\providecommand \href [0]{\begingroup \@sanitize@url \@href}%
\providecommand \@href[1]{\@@startlink{#1}\@@href}%
\providecommand \@@href[1]{\endgroup#1\@@endlink}%
\providecommand \@sanitize@url [0]{\catcode `\\12\catcode `\$12\catcode
  `\&12\catcode `\#12\catcode `\^12\catcode `\_12\catcode `\%12\relax}%
\providecommand \@@startlink[1]{}%
\providecommand \@@endlink[0]{}%
\providecommand \url  [0]{\begingroup\@sanitize@url \@url }%
\providecommand \@url [1]{\endgroup\@href {#1}{\urlprefix }}%
\providecommand \urlprefix  [0]{URL }%
\providecommand \Eprint [0]{\href }%
\providecommand \doibase [0]{https://doi.org/}%
\providecommand \selectlanguage [0]{\@gobble}%
\providecommand \bibinfo  [0]{\@secondoftwo}%
\providecommand \bibfield  [0]{\@secondoftwo}%
\providecommand \translation [1]{[#1]}%
\providecommand \BibitemOpen [0]{}%
\providecommand \bibitemStop [0]{}%
\providecommand \bibitemNoStop [0]{.\EOS\space}%
\providecommand \EOS [0]{\spacefactor3000\relax}%
\providecommand \BibitemShut  [1]{\csname bibitem#1\endcsname}%
\let\auto@bib@innerbib\@empty
\bibitem [{\citenamefont {James}(2018)}]{AdvEngMath}%
  \BibitemOpen
  \bibinfo {editor} {\bibfnamefont {G.}~\bibnamefont {James}},\ ed.,\
  \href@noop {} {\emph {\bibinfo {title} {Advanced Modern Engineering
  Mathematics}}},\ \bibinfo {edition} {fifth edition}\ ed.\ (\bibinfo
  {publisher} {{Pearson Education}},\ \bibinfo {address} {{Harlow, United
  Kingdom}},\ \bibinfo {year} {2018})\BibitemShut {NoStop}%
\bibitem [{\citenamefont {Oppenheim}\ \emph {et~al.}(1983)\citenamefont
  {Oppenheim}, \citenamefont {Willsky},\ and\ \citenamefont
  {Young}}]{oppenheimSignalsSystems1983}%
  \BibitemOpen
  \bibfield  {author} {\bibinfo {author} {\bibfnamefont {A.~V.}\ \bibnamefont
  {Oppenheim}}, \bibinfo {author} {\bibfnamefont {A.~S.}\ \bibnamefont
  {Willsky}},\ and\ \bibinfo {author} {\bibfnamefont {I.~T.}\ \bibnamefont
  {Young}},\ }\href@noop {} {\emph {\bibinfo {title} {Signals and Systems}}},\
  Prentice-{{Hall}} Signal Processing Series\ (\bibinfo  {publisher}
  {{Prentice-Hall}},\ \bibinfo {address} {{Englewood Cliffs, NJ}},\ \bibinfo
  {year} {1983})\BibitemShut {NoStop}%
\bibitem [{\citenamefont {Ogata}(2010)}]{ogataModernControlEngineering2010}%
  \BibitemOpen
  \bibfield  {author} {\bibinfo {author} {\bibfnamefont {K.}~\bibnamefont
  {Ogata}},\ }\href@noop {} {\emph {\bibinfo {title} {Modern Control
  Engineering}}},\ \bibinfo {edition} {5th}\ ed.,\ Prentice-{{Hall}} Electrical
  Engineering Series. {{Instrumentation}} and Controls Series\ (\bibinfo
  {publisher} {{Prentice-Hall}},\ \bibinfo {address} {{Boston}},\ \bibinfo
  {year} {2010})\BibitemShut {NoStop}%
\bibitem [{\citenamefont {Debnath}\ and\ \citenamefont
  {Bhatta}(2007)}]{debnathIntegralTransformsTheir2007}%
  \BibitemOpen
  \bibfield  {author} {\bibinfo {author} {\bibfnamefont {L.}~\bibnamefont
  {Debnath}}\ and\ \bibinfo {author} {\bibfnamefont {D.}~\bibnamefont
  {Bhatta}},\ }\href@noop {} {\emph {\bibinfo {title} {Integral Transforms and
  Their Applications}}},\ \bibinfo {edition} {2nd}\ ed.\ (\bibinfo  {publisher}
  {{Chapman \& Hall/CRC}},\ \bibinfo {address} {{Boca Raton, Fla.}},\ \bibinfo
  {year} {2007})\BibitemShut {NoStop}%
\bibitem [{\citenamefont {Moore}\ and\ \citenamefont
  {Mertens}(2011)}]{natureofcomputation}%
  \BibitemOpen
  \bibfield  {author} {\bibinfo {author} {\bibfnamefont {C.}~\bibnamefont
  {Moore}}\ and\ \bibinfo {author} {\bibfnamefont {S.}~\bibnamefont
  {Mertens}},\ }\href@noop {} {\emph {\bibinfo {title} {The Nature of
  Computation}}}\ (\bibinfo  {publisher} {{Oxford University Press}},\ \bibinfo
  {address} {{Oxford}},\ \bibinfo {year} {2011})\BibitemShut {NoStop}%
\bibitem [{\citenamefont {Martins}\ and\ \citenamefont
  {Lambe}(2013)}]{martinsMultidisciplinaryDesignOptimization2013}%
  \BibitemOpen
  \bibfield  {author} {\bibinfo {author} {\bibfnamefont {J.~R. R.~A.}\
  \bibnamefont {Martins}}\ and\ \bibinfo {author} {\bibfnamefont {A.~B.}\
  \bibnamefont {Lambe}},\ }\bibfield  {title} {\bibinfo {title}
  {Multidisciplinary {{Design Optimization}}: {{A Survey}} of
  {{Architectures}}},\ }\href {https://doi.org/10.2514/1.J051895} {\bibfield
  {journal} {\bibinfo  {journal} {AIAA Journal}\ }\textbf {\bibinfo {volume}
  {51}},\ \bibinfo {pages} {2049} (\bibinfo {year} {2013})}\BibitemShut
  {NoStop}%
\bibitem [{\citenamefont {Polak}(1997)}]{PolakOptimization}%
  \BibitemOpen
  \bibfield  {author} {\bibinfo {author} {\bibfnamefont {E.}~\bibnamefont
  {Polak}},\ }\href {https://doi.org/10.1007/978-1-4612-0663-7} {\emph
  {\bibinfo {title} {Optimization}}},\ edited by\ \bibinfo {editor}
  {\bibfnamefont {J.~E.}\ \bibnamefont {Marsden}}, \bibinfo {editor}
  {\bibfnamefont {L.}~\bibnamefont {Sirovich}},\ and\ \bibinfo {editor}
  {\bibfnamefont {F.}~\bibnamefont {John}},\ \bibinfo {series} {Applied
  {{Mathematical Sciences}}}, Vol.\ \bibinfo {volume} {124}\ (\bibinfo
  {publisher} {{Springer New York}},\ \bibinfo {address} {{New York, NY}},\
  \bibinfo {year} {1997})\BibitemShut {NoStop}%
\bibitem [{\citenamefont {Shannon}(1948)}]{shannon}%
  \BibitemOpen
  \bibfield  {author} {\bibinfo {author} {\bibfnamefont {C.}~\bibnamefont
  {Shannon}},\ }\bibfield  {title} {\bibinfo {title} {A mathematical theory of
  communication},\ }\href {https://doi.org/10.1002/j.1538-7305.1948.tb01338.x}
  {\bibfield  {journal} {\bibinfo  {journal} {Bell Syst. Tech. J.}\ }\textbf
  {\bibinfo {volume} {27}},\ \bibinfo {pages} {379} (\bibinfo {year}
  {1948})}\BibitemShut {NoStop}%
\bibitem [{\citenamefont {Sethna}(2021)}]{Sethna2021}%
  \BibitemOpen
  \bibfield  {author} {\bibinfo {author} {\bibfnamefont {J.}~\bibnamefont
  {Sethna}},\ }\href@noop {} {\emph {\bibinfo {title} {Statistical Mechanics:
  Entropy, Order Parameters, and Complexity}}}\ (\bibinfo  {publisher} {{Oxford
  University Press, USA}},\ \bibinfo {year} {2021})\BibitemShut {NoStop}%
\bibitem [{\citenamefont {Kirkpatrick}\ \emph {et~al.}(1983)\citenamefont
  {Kirkpatrick}, \citenamefont {Gelatt},\ and\ \citenamefont
  {Vecchi}}]{kirkpatrick1983}%
  \BibitemOpen
  \bibfield  {author} {\bibinfo {author} {\bibfnamefont {S.}~\bibnamefont
  {Kirkpatrick}}, \bibinfo {author} {\bibfnamefont {C.~D.}\ \bibnamefont
  {Gelatt}},\ and\ \bibinfo {author} {\bibfnamefont {M.~P.}\ \bibnamefont
  {Vecchi}},\ }\bibfield  {title} {\bibinfo {title} {Optimization by simulated
  annealing},\ }\href {https://doi.org/10.1126/science.220.4598.671} {\bibfield
   {journal} {\bibinfo  {journal} {Science}\ }\textbf {\bibinfo {volume}
  {220}},\ \bibinfo {pages} {671} (\bibinfo {year} {1983})}\BibitemShut
  {NoStop}%
\bibitem [{\citenamefont {{van Anders}}\ \emph {et~al.}(2015)\citenamefont
  {{van Anders}}, \citenamefont {Klotsa}, \citenamefont {Karas}, \citenamefont
  {Dodd},\ and\ \citenamefont {Glotzer}}]{digitalalchemy}%
  \BibitemOpen
  \bibfield  {author} {\bibinfo {author} {\bibfnamefont {G.}~\bibnamefont {{van
  Anders}}}, \bibinfo {author} {\bibfnamefont {D.}~\bibnamefont {Klotsa}},
  \bibinfo {author} {\bibfnamefont {A.~S.}\ \bibnamefont {Karas}}, \bibinfo
  {author} {\bibfnamefont {P.~M.}\ \bibnamefont {Dodd}},\ and\ \bibinfo
  {author} {\bibfnamefont {S.~C.}\ \bibnamefont {Glotzer}},\ }\bibfield
  {title} {\bibinfo {title} {Digital {{Alchemy}} for {{Materials Design}}:
  {{Colloids}} and {{Beyond}}},\ }\href
  {https://doi.org/10.1021/acsnano.5b04181} {\bibfield  {journal} {\bibinfo
  {journal} {ACS Nano}\ }\textbf {\bibinfo {volume} {9}},\ \bibinfo {pages}
  {9542} (\bibinfo {year} {2015})}\BibitemShut {NoStop}%
\bibitem [{\citenamefont {Miskin}\ \emph {et~al.}(2016)\citenamefont {Miskin},
  \citenamefont {Khaira}, \citenamefont {{de Pablo}},\ and\ \citenamefont
  {Jaeger}}]{miskinpnas}%
  \BibitemOpen
  \bibfield  {author} {\bibinfo {author} {\bibfnamefont {M.~Z.}\ \bibnamefont
  {Miskin}}, \bibinfo {author} {\bibfnamefont {G.}~\bibnamefont {Khaira}},
  \bibinfo {author} {\bibfnamefont {J.~J.}\ \bibnamefont {{de Pablo}}},\ and\
  \bibinfo {author} {\bibfnamefont {H.~M.}\ \bibnamefont {Jaeger}},\ }\bibfield
   {title} {\bibinfo {title} {Turning statistical physics models into materials
  design engines},\ }\href {https://doi.org/10.1073/pnas.1509316112} {\bibfield
   {journal} {\bibinfo  {journal} {Proc. Natl. Acad. Sci. U.S.A.}\ }\textbf
  {\bibinfo {volume} {113}},\ \bibinfo {pages} {34} (\bibinfo {year}
  {2016})}\BibitemShut {NoStop}%
\bibitem [{\citenamefont {Chitnelawong}\ \emph {et~al.}(2023)\citenamefont
  {Chitnelawong}, \citenamefont {Klishin}, \citenamefont {MacKay},
  \citenamefont {Singer},\ and\ \citenamefont {{van Anders}}}]{nfl}%
  \BibitemOpen
  \bibfield  {author} {\bibinfo {author} {\bibfnamefont {P.}~\bibnamefont
  {Chitnelawong}}, \bibinfo {author} {\bibfnamefont {A.~A.}\ \bibnamefont
  {Klishin}}, \bibinfo {author} {\bibfnamefont {N.}~\bibnamefont {MacKay}},
  \bibinfo {author} {\bibfnamefont {D.~J.}\ \bibnamefont {Singer}},\ and\
  \bibinfo {author} {\bibfnamefont {G.}~\bibnamefont {{van Anders}}},\
  }\href@noop {} {\bibinfo {title} {No {{Free Lunch}} for {{Avoiding Clustering
  Vulnerabilities}} in {{Distributed Systems}}}} (\bibinfo {year} {2023}),\
  \Eprint {https://arxiv.org/abs/2308.05196} {arxiv:2308.05196 [cond-mat,
  physics:physics]} \BibitemShut {NoStop}%
\bibitem [{\citenamefont {Aliahmadi}\ \emph {et~al.}(2023)\citenamefont
  {Aliahmadi}, \citenamefont {Beckett}, \citenamefont {Connolly}, \citenamefont
  {Chen},\ and\ \citenamefont {{van Anders}}}]{Flashpoints}%
  \BibitemOpen
  \bibfield  {author} {\bibinfo {author} {\bibfnamefont {H.}~\bibnamefont
  {Aliahmadi}}, \bibinfo {author} {\bibfnamefont {M.}~\bibnamefont {Beckett}},
  \bibinfo {author} {\bibfnamefont {S.}~\bibnamefont {Connolly}}, \bibinfo
  {author} {\bibfnamefont {D.}~\bibnamefont {Chen}},\ and\ \bibinfo {author}
  {\bibfnamefont {G.}~\bibnamefont {{van Anders}}},\ }\href@noop {} {\bibinfo
  {title} {Flashpoints {{Signal Hidden Inherent Instabilities}} in {{Land-Use
  Planning}}}} (\bibinfo {year} {2023}),\ \Eprint
  {https://arxiv.org/abs/2308.07714} {arxiv:2308.07714 [cond-mat,
  physics:physics]} \BibitemShut {NoStop}%
\bibitem [{\citenamefont {Cabrera}\ \emph {et~al.}(2023)\citenamefont
  {Cabrera}, \citenamefont {Babayan}, \citenamefont {Aliahmadi}, \citenamefont
  {Chen},\ and\ \citenamefont {{van Anders}}}]{slo-go}%
  \BibitemOpen
  \bibfield  {author} {\bibinfo {author} {\bibfnamefont {S.}~\bibnamefont
  {Cabrera}}, \bibinfo {author} {\bibfnamefont {I.}~\bibnamefont {Babayan}},
  \bibinfo {author} {\bibfnamefont {H.}~\bibnamefont {Aliahmadi}}, \bibinfo
  {author} {\bibfnamefont {D.}~\bibnamefont {Chen}},\ and\ \bibinfo {author}
  {\bibfnamefont {G.}~\bibnamefont {{van Anders}}},\ }\href@noop {} {\bibinfo
  {title} {{{SLO}}/{{GO Degradation-Loss Sensitivity}} in {{Climate-Human
  System Coupling}}}} (\bibinfo {year} {2023}),\ \Eprint
  {https://arxiv.org/abs/2311.17905} {arxiv:2311.17905 [physics]} \BibitemShut
  {NoStop}%
\bibitem [{\citenamefont {Geng}\ \emph {et~al.}(2019)\citenamefont {Geng},
  \citenamefont {{van Anders}}, \citenamefont {Dodd}, \citenamefont
  {Dshemuchadse},\ and\ \citenamefont {Glotzer}}]{engent}%
  \BibitemOpen
  \bibfield  {author} {\bibinfo {author} {\bibfnamefont {Y.}~\bibnamefont
  {Geng}}, \bibinfo {author} {\bibfnamefont {G.}~\bibnamefont {{van Anders}}},
  \bibinfo {author} {\bibfnamefont {P.~M.}\ \bibnamefont {Dodd}}, \bibinfo
  {author} {\bibfnamefont {J.}~\bibnamefont {Dshemuchadse}},\ and\ \bibinfo
  {author} {\bibfnamefont {S.~C.}\ \bibnamefont {Glotzer}},\ }\bibfield
  {title} {\bibinfo {title} {Engineering {{Entropy}} for the {{Inverse Design}}
  of {{Colloidal Crystals}} from {{Hard Shapes}}},\ }\href
  {https://doi.org/10.1126/sciadv.aaw0514} {\bibfield  {journal} {\bibinfo
  {journal} {Science Advances}\ }\textbf {\bibinfo {volume} {5}},\ \bibinfo
  {pages} {eaaw0514} (\bibinfo {year} {2019})}\BibitemShut {NoStop}%
\bibitem [{\citenamefont {Gould}\ \emph {et~al.}(2023)\citenamefont {Gould},
  \citenamefont {Tobochnik},\ and\ \citenamefont
  {Christian}}]{GouldTobochnik3e}%
  \BibitemOpen
  \bibfield  {author} {\bibinfo {author} {\bibfnamefont {H.}~\bibnamefont
  {Gould}}, \bibinfo {author} {\bibfnamefont {J.}~\bibnamefont {Tobochnik}},\
  and\ \bibinfo {author} {\bibfnamefont {W.}~\bibnamefont {Christian}},\
  }\href@noop {} {\emph {\bibinfo {title} {An Introduction to Computer
  Simulation Methods: Applications to Physical Systems}}},\ \bibinfo {edition}
  {revised third edition}\ ed.\ (\bibinfo  {publisher} {{Amazon Fulfillment}},\
  \bibinfo {address} {{Wroc{\l}aw}},\ \bibinfo {year} {2023})\BibitemShut
  {NoStop}%
\bibitem [{\citenamefont {Cersonsky}\ \emph {et~al.}(2018)\citenamefont
  {Cersonsky}, \citenamefont {{van Anders}}, \citenamefont {Dodd},\ and\
  \citenamefont {Glotzer}}]{packingassembly}%
  \BibitemOpen
  \bibfield  {author} {\bibinfo {author} {\bibfnamefont {R.}~\bibnamefont
  {Cersonsky}}, \bibinfo {author} {\bibfnamefont {G.}~\bibnamefont {{van
  Anders}}}, \bibinfo {author} {\bibfnamefont {P.~M.}\ \bibnamefont {Dodd}},\
  and\ \bibinfo {author} {\bibfnamefont {S.~C.}\ \bibnamefont {Glotzer}},\
  }\bibfield  {title} {\bibinfo {title} {Relevance of packing to colloidal
  self-assembly},\ }\href {https://doi.org/10.1073/pnas.1720139115} {\bibfield
  {journal} {\bibinfo  {journal} {Proc. Natl. Acad. Sci. USA}\ }\textbf
  {\bibinfo {volume} {115}},\ \bibinfo {pages} {1439} (\bibinfo {year}
  {2018})},\ \Eprint {https://arxiv.org/abs/1712.02473} {arxiv:1712.02473
  [cond-mat.soft]} \BibitemShut {NoStop}%
\bibitem [{\citenamefont {Zhou}\ \emph {et~al.}(2019)\citenamefont {Zhou},
  \citenamefont {Proctor}, \citenamefont {{van Anders}},\ and\ \citenamefont
  {Glotzer}}]{Alch-MD}%
  \BibitemOpen
  \bibfield  {author} {\bibinfo {author} {\bibfnamefont {P.}~\bibnamefont
  {Zhou}}, \bibinfo {author} {\bibfnamefont {J.}~\bibnamefont {Proctor}},
  \bibinfo {author} {\bibfnamefont {G.}~\bibnamefont {{van Anders}}},\ and\
  \bibinfo {author} {\bibfnamefont {S.~C.}\ \bibnamefont {Glotzer}},\
  }\bibfield  {title} {\bibinfo {title} {Alchemical molecular dynamics for
  inverse design},\ }\href {https://doi.org/10.1080/00268976.2019.1680886}
  {\bibfield  {journal} {\bibinfo  {journal} {Molecular Physics}\ }\textbf
  {\bibinfo {volume} {117}},\ \bibinfo {pages} {3968} (\bibinfo {year}
  {2019})}\BibitemShut {NoStop}%
\bibitem [{\citenamefont {Du}\ \emph {et~al.}(2020)\citenamefont {Du},
  \citenamefont {{van Anders}}, \citenamefont {Dshemuchadse},\ and\
  \citenamefont {Glotzer}}]{bccfccdesign}%
  \BibitemOpen
  \bibfield  {author} {\bibinfo {author} {\bibfnamefont {C.~X.}\ \bibnamefont
  {Du}}, \bibinfo {author} {\bibfnamefont {G.}~\bibnamefont {{van Anders}}},
  \bibinfo {author} {\bibfnamefont {J.}~\bibnamefont {Dshemuchadse}},\ and\
  \bibinfo {author} {\bibfnamefont {S.~C.}\ \bibnamefont {Glotzer}},\
  }\bibfield  {title} {\bibinfo {title} {{Inverse design of compression-induced
  Solid--Solid transitions in colloids}},\ }\href
  {https://doi.org/10.1080/08927022.2020.1798005} {\bibfield  {journal}
  {\bibinfo  {journal} {Molecular Simulation}\ }\textbf {\bibinfo {volume}
  {46}},\ \bibinfo {pages} {1037} (\bibinfo {year} {2020})}\BibitemShut
  {NoStop}%
\bibitem [{\citenamefont {Teich}\ \emph {et~al.}(2019)\citenamefont {Teich},
  \citenamefont {{van Anders}},\ and\ \citenamefont
  {Glotzer}}]{identitycrisis}%
  \BibitemOpen
  \bibfield  {author} {\bibinfo {author} {\bibfnamefont {E.~G.}\ \bibnamefont
  {Teich}}, \bibinfo {author} {\bibfnamefont {G.}~\bibnamefont {{van
  Anders}}},\ and\ \bibinfo {author} {\bibfnamefont {S.~C.}\ \bibnamefont
  {Glotzer}},\ }\bibfield  {title} {\bibinfo {title} {Identity crisis in
  alchemical space drives the entropic colloidal glass transition},\ }\href
  {https://doi.org/10.1038/s41467-018-07977-2} {\bibfield  {journal} {\bibinfo
  {journal} {Nat. Commun.}\ }\textbf {\bibinfo {volume} {10}},\ \bibinfo
  {pages} {64} (\bibinfo {year} {2019})}\BibitemShut {NoStop}%
\bibitem [{\citenamefont {Klishin}\ \emph {et~al.}(2018)\citenamefont
  {Klishin}, \citenamefont {Shields}, \citenamefont {Singer},\ and\
  \citenamefont {{van Anders}}}]{systemphys}%
  \BibitemOpen
  \bibfield  {author} {\bibinfo {author} {\bibfnamefont {A.~A.}\ \bibnamefont
  {Klishin}}, \bibinfo {author} {\bibfnamefont {C.~P.}\ \bibnamefont
  {Shields}}, \bibinfo {author} {\bibfnamefont {D.~J.}\ \bibnamefont
  {Singer}},\ and\ \bibinfo {author} {\bibfnamefont {G.}~\bibnamefont {{van
  Anders}}},\ }\bibfield  {title} {\bibinfo {title} {Statistical physics of
  design},\ }\href {https://doi.org/10.1088/1367-2630/aae72a} {\bibfield
  {journal} {\bibinfo  {journal} {New J. Phys.}\ }\textbf {\bibinfo {volume}
  {20}},\ \bibinfo {pages} {103038} (\bibinfo {year} {2018})},\ \Eprint
  {https://arxiv.org/abs/1709.03388} {arxiv:1709.03388 [physics.soc-ph]}
  \BibitemShut {NoStop}%
\bibitem [{\citenamefont {Klishin}\ \emph {et~al.}(2020)\citenamefont
  {Klishin}, \citenamefont {Kirkley}, \citenamefont {Singer},\ and\
  \citenamefont {{van Anders}}}]{robustdesign}%
  \BibitemOpen
  \bibfield  {author} {\bibinfo {author} {\bibfnamefont {A.~A.}\ \bibnamefont
  {Klishin}}, \bibinfo {author} {\bibfnamefont {A.}~\bibnamefont {Kirkley}},
  \bibinfo {author} {\bibfnamefont {D.~J.}\ \bibnamefont {Singer}},\ and\
  \bibinfo {author} {\bibfnamefont {G.}~\bibnamefont {{van Anders}}},\
  }\bibfield  {title} {\bibinfo {title} {Robust design from systems physics},\
  }\href {https://doi.org/10.1038/s41598-020-70980-5} {\bibfield  {journal}
  {\bibinfo  {journal} {Scientific Reports}\ }\textbf {\bibinfo {volume}
  {10}},\ \bibinfo {pages} {14334} (\bibinfo {year} {2020})},\ \Eprint
  {https://arxiv.org/abs/1805.02691} {arxiv:1805.02691 [physics.soc-ph]}
  \BibitemShut {NoStop}%
\bibitem [{\citenamefont {Klishin}\ \emph {et~al.}(2021)\citenamefont
  {Klishin}, \citenamefont {Singer},\ and\ \citenamefont {{van Anders}}}]{aaa}%
  \BibitemOpen
  \bibfield  {author} {\bibinfo {author} {\bibfnamefont {A.~A.}\ \bibnamefont
  {Klishin}}, \bibinfo {author} {\bibfnamefont {D.~J.}\ \bibnamefont
  {Singer}},\ and\ \bibinfo {author} {\bibfnamefont {G.}~\bibnamefont {{van
  Anders}}},\ }\bibfield  {title} {\bibinfo {title} {Avoidance, adjacency, and
  association in distributed system design},\ }\href
  {https://doi.org/10.1088/2632-072X/abe27f} {\bibfield  {journal} {\bibinfo
  {journal} {J. Phys. Complexity}\ }\textbf {\bibinfo {volume} {2}},\ \bibinfo
  {pages} {025015} (\bibinfo {year} {2021})},\ \Eprint
  {https://arxiv.org/abs/2010.00141} {arxiv:2010.00141 [physics.soc-ph]}
  \BibitemShut {NoStop}%
\bibitem [{\citenamefont {Jaynes}(1957)}]{jaynes1}%
  \BibitemOpen
  \bibfield  {author} {\bibinfo {author} {\bibfnamefont {E.~T.}\ \bibnamefont
  {Jaynes}},\ }\bibfield  {title} {\bibinfo {title} {Information theory and
  statistical mechanics},\ }\href {https://doi.org/10.1103/PhysRev.106.620}
  {\bibfield  {journal} {\bibinfo  {journal} {Phys. Rev.}\ }\textbf {\bibinfo
  {volume} {106}},\ \bibinfo {pages} {620} (\bibinfo {year}
  {1957})}\BibitemShut {NoStop}%
\bibitem [{\citenamefont {Frenkel}\ and\ \citenamefont {Smit}(2002)}]{frenkel}%
  \BibitemOpen
  \bibfield  {author} {\bibinfo {author} {\bibfnamefont {D.}~\bibnamefont
  {Frenkel}}\ and\ \bibinfo {author} {\bibfnamefont {B.}~\bibnamefont {Smit}},\
  }\href@noop {} {\emph {\bibinfo {title} {Understanding Molecular Simulation;
  from Algorithms to Applications}}}\ (\bibinfo  {publisher} {{Academic
  Press}},\ \bibinfo {year} {2002})\BibitemShut {NoStop}%
\bibitem [{\citenamefont {Aliahmadi}\ \emph {et~al.}(2024)\citenamefont
  {Aliahmadi}, \citenamefont {Sheedy}, \citenamefont {Perez},\ and\
  \citenamefont {Van~Anders}}]{HyperoptMorphogenesis}%
  \BibitemOpen
  \bibfield  {author} {\bibinfo {author} {\bibfnamefont {H.}~\bibnamefont
  {Aliahmadi}}, \bibinfo {author} {\bibfnamefont {A.}~\bibnamefont {Sheedy}},
  \bibinfo {author} {\bibfnamefont {R.}~\bibnamefont {Perez}},\ and\ \bibinfo
  {author} {\bibfnamefont {G.}~\bibnamefont {Van~Anders}},\ }\bibfield  {title}
  {\bibinfo {title} {Hyperoptimization insight for computational
  morphogenesis},\ }\href@noop {} {\  (\bibinfo {year} {2024})}\BibitemShut
  {NoStop}%
\bibitem [{\citenamefont {Landau}\ and\ \citenamefont {Lifshitz}(1980)}]{LLv5}%
  \BibitemOpen
  \bibfield  {author} {\bibinfo {author} {\bibfnamefont {L.~D.}\ \bibnamefont
  {Landau}}\ and\ \bibinfo {author} {\bibfnamefont {E.~M.}\ \bibnamefont
  {Lifshitz}},\ }\href@noop {} {\emph {\bibinfo {title} {Statistical Physics,
  Part 1}}},\ \bibinfo {edition} {3rd}\ ed.\ (\bibinfo  {publisher}
  {{Butterworth-Heinemann}},\ \bibinfo {address} {{Oxford}},\ \bibinfo {year}
  {1980})\BibitemShut {NoStop}%
\bibitem [{\citenamefont {Goldenfeld}(1992)}]{goldenfeld}%
  \BibitemOpen
  \bibfield  {author} {\bibinfo {author} {\bibfnamefont {N.}~\bibnamefont
  {Goldenfeld}},\ }\href@noop {} {\emph {\bibinfo {title} {Lectures on Phase
  Transitions and the Renormalization Group}}}\ (\bibinfo  {publisher}
  {{Addison-Wesley}},\ \bibinfo {address} {{Reading MA}},\ \bibinfo {year}
  {1992})\BibitemShut {NoStop}%
\bibitem [{\citenamefont {{Zinn-Justin}}(2002)}]{zinnjustin}%
  \BibitemOpen
  \bibfield  {author} {\bibinfo {author} {\bibfnamefont {J.}~\bibnamefont
  {{Zinn-Justin}}},\ }\href@noop {} {\emph {\bibinfo {title} {Quantum Field
  Theory and Critical Phenomena}}},\ \bibinfo {edition} {4th}\ ed.\ (\bibinfo
  {publisher} {{Oxford University Press}},\ \bibinfo {address} {{Oxford}},\
  \bibinfo {year} {2002})\BibitemShut {NoStop}%
\end{thebibliography}
\end{document}